\documentclass[preprint,12pt]{elsarticle}

\usepackage{amssymb}
\usepackage{amsmath,amsfonts}
\usepackage{algorithmic}
\usepackage{algorithm}
\usepackage{array}
\usepackage[caption=false,font=normalsize,labelfont=sf,textfont=sf]{subfig}
\usepackage{textcomp}
\usepackage{stfloats}
\usepackage{url}
\usepackage{verbatim}
\usepackage{graphicx}
\usepackage{xcolor}
\usepackage{changepage}

\journal{Vehicular Communications}

\begin{document}

\begin{frontmatter}



\title{Blockchain Model for Environment/Infrastructure Monitoring in Cloud-Enabled High-Altitude Platform Systems}


\author[inst1]{Khaleel Mershad\corref{cor1}}
\ead{khaleel.mershad@lau.edu.lb}
\affiliation[inst1]{organization={Computer Science and Mathematics Department},
            addressline={Lebanese American University (LAU)}, 
            city={Beirut},
            country={Lebanon}}

\author[inst2]{Hayssam Dahrouj}

\affiliation[inst2]{organization={Electrical Engineering Department},
            addressline={University of Sharjah (UoS)}, 
            city={Sharjah},
            country={United Arab Emirates}}

\cortext[cor1]{corresponding author}

\begin{abstract}
The recently accentuated features of augmenting conventional wireless networks with high altitude platform systems (HAPS) have fueled a plethora of applications, which promise to offer new services to ground users, as well to enhance the efficiency and pervasion of existing applications. Cloud-enabled HAPS, which aims to create HAPS-based datacenters that offer cloud services to users, has particularly emerged as a promising key enabler to provide large-scale equitable services from the sky. Although offering cloud services from the HAPS proves to be efficient, its practical deployment at the stratosphere level still faces many challenges such as high energy requirements, physical maintenance, and is particularly prone to security considerations. Safeguarding the cloud-enabled HAPS against various cyberattacks is a necessity to guarantee its safe operation. 
This paper proposes a blockchain model to secure cloud-enabled HAPS networks that contain a large number of HAPS stations from recurring cyberattacks within the context of the environment and infrastructure monitoring (EIM) application. 
To this end, the paper first presents a detailed blockchain framework, and describes the ways of integrating the developed framework into the various system components. We then discuss the details of the system implementation, including the storing and consuming of cloud transactions, the generation of new blocks, and the blockchain consensus protocol that is tailored to the EIM requirements. Finally, we present numerical simulations that illustrate the performance of the system in terms of throughput, latency, and resilience to attacks. 
\end{abstract} 




\end{frontmatter}


\section{Introduction}

The premises of High Altitude Platform Systems (HAPS) at enhancing ground-level networks have developed drastically over the recent years. In HAPS platforms, a HAPS station is deployed in the stratosphere at altitudes in the order of 18-20 km to enhance the terrestrial network architecture and serve ground users through providing wireless coverage, backhauling small and isolated base stations (BSs), supporting Internet of Things (IoT) applications, assisting intelligent transportation systems (ITS), and handling LEO satellite handoffs \cite{kurt2021vision}. {\color{black}{HAPS networks have several advantages that distinguish them from other aerial networks such as satellite and UAV networks, e.g.,  \cite{Zhang_ICC2020,Zhang_TVT2023,Zhang_IoT2023}.}} These advantages include enhanced quality-of-service in dense areas, quasi-stationary deployment for robust sky-level communication, reduced round-trip delay, and an expansive wireless footprint \cite{kurt2021vision}.

Among the key-enablers which are expected to proliferate HAPS applications is the cloud-enabled HAPS (C-HAPS) \cite{mershad2021cloud}. C-HAPS allows the HAPS stations to be utilized as flying cloud datacenters that offer cloud services from the sky. Such integration of physical and cloud services improves the cloud scalability and enhances the quality, speed, and range of the offered services for ground users. Further, the HAPS strategic position enables cloud providers to reach out to a larger range of customers. 
Moreover, augmenting the HAPS with cloud-computing capabilities allows to better serve remote and disconnected areas either directly or via ground/aerial gateways \cite{alam2021high} by means of implementing interference management schemes at the cloud level. \color{black}{Such powerful functionalities of cloud-enabled HAPS, in fact, spearhead a handful of cloud-computing applications from the sky, thereby increasing the system storage capacity, improving data processing, minimizing the cost savings, and increasing the system scalability. In an integrated large-scale air-ground system, the HAPS can also be viewed as an edge or a fog layer, which serves to better extend the system connectivity and computational capabilities.}\color{black}

Despite the multiple prospects of C-HAPS, several challenges need to be addressed in order to successfully realize the integration of cloud services into HAPS \cite{mershad2021cloud}. Among such challenges is the system security issues, especially those related to HAPS communications, data, and programs. In general, the security of cloud services is one of the most important factors to ensure their success and continuity. When offered from the HAPS, cloud services need to be carefully designed to prevent attackers from eavesdropping on the connections between the HAPS station and the ground/aerial nodes in attempts to access private data. Equally important is to store the cloud data within the HAPS station in a secure fashion. To this end, this paper proposes a blockchain model as a possible solution for securing the C-HAPS operations. Utilizing the blockchain offers several benefits to cloud platforms, such as data immutability, built-in cryptography, and distributed management. Moreover, the blockchain can safeguard the HAPS services from major cyberattacks such as Man-in-the-Middle, Denial of Service (DoS), unauthorized access, etc. 

The model considered in this paper focuses on one specific timely application of C-HAPS, namely the one related to environment and infrastructure monitoring (EIM). The EIM application is usually deployed within a large-scale network that contains multiple wireless sensor networks (WSN) comprising a huge number of sensor nodes. Examples of such networks include forests' WSNs, smart grid WSNs, maritime WSNs, etc. In such networks, several HAPS stations are used to connect to the WSNs via ground sinks/gateways that are equipped with HAPS communication modules. Each HAPS station can be standalone or part of a mini-HAPS constellation. The IoT sensors within a WSN are clustered such that each cluster head (CH) collects the readings of the sensors in the cluster and sends them to the sink/gateway, which in turn forwards them to the nearest HAPS station. Within the HAPS datacenter, the sensors' readings are analyzed and processed to produce results that are consumed by the cloud users. Among the many premises of the EIM application is that it is delay-sensitive, as it deals with emergencies and situations that require fast reaction. \color{black}Current cloud systems that depend on terrestrial networks fail to satisfy the strict latency requirements of the EIM application while providing strong security mechanisms that are able to detect and prevent cyberattacks. Hence, the motivation behind the proposed system is twofold, mainly to strengthen the security of the EIM cloud service via the blockchain, and to achieve acceptable latency by modifying the blockchain architecture and consensus model. The simulations performed in Section \ref{Sec_performance} show that our proposed model reduces the latency of data transactions from 1 sec to 200 ms and is able to detect attacks with an average of 85\% as compared to 65\% that is achieved by a ground-based system. 

\color{black}The contributions of this paper can be summarized as follows:
\begin{itemize}
\item 
We study the implementation of a large-scale EIM application within the context of C-HAPS. Current systems that implement the EIM utilize the terrestrial networks and cloud datacenters \cite{lombardo2017wireless, ouni2022framework}, with some systems exploiting unmanned aerial vehicle (UAV) networks to assist in the monitoring operations \cite{pan2021uav}. In the proposed model, the EIM application is deployed within the C-HAPS to make use of the reduced delay characteristic of HAPS. In the simulations, we show the efficiency of this method as compared to the traditional ground implementation. \color{black}Based on the simulation results, our proposed system reduces the data transaction latency from an average of 1 sec in ground-based systems to 200 ms, which is a crucial improvement that plays a major role in the success of a large-scale EIM application.\color{black}
\item 
We design and implement a blockchain model for securing the C-HAPS EIM application. The proposed model makes use of the powerful and secure HAPS stations to generate the blockchain blocks, and of ground/aerial gateway stations to validate the blocks and strengthen the blockchain consensus protocol. This method reduces the time needed to add a transaction to the blockchain and satisfies the delay requirements of the EIM application. \color{black}In addition, the proposed system is much more resistant to attacks as compared to a system that does not utilize the blockchain. From the simulation results, the rate of malicious node detection is higher by 20\% (on average) in our system as compared to a traditional cloud system. To the best of our knowledge, this is the first paper of its kind that studies the blockchain as a secure and efficient solution for the C-HAPS EIM application. We prove that the introduction of the blockchain highly strengthens the security of the cloud service. \color{black} 
\item 
We validate the proposed system by testing its performance using the NS-3 network simulation software. The results illustrate the superiority of our model in terms of blockchain throughput, transaction latency, blockchain consensus time, and attack detection rate, as compared to a ground-based similar system. 
\end{itemize}

The remaining of this paper is organized as follows: In the next section, we review the background of C-HAPS and blockchain. Section \ref{Sec_review} summarizes the state of the art related to the EIM application. Section \ref{Sec_proposed} describes the details of the proposed blockchain system, its integration into the C-HAPS environment, and its transaction management and consensus operations. In Section \ref{Sec_performance}, we present the simulations that we performed to test the proposed system using the network simulator 3 tool. Finally, Section \ref{Sec_conc} concludes the paper by shedding light on future work and potential enhancements.

\begin{figure*}[!t]
\centering
\includegraphics[width=5.6in]{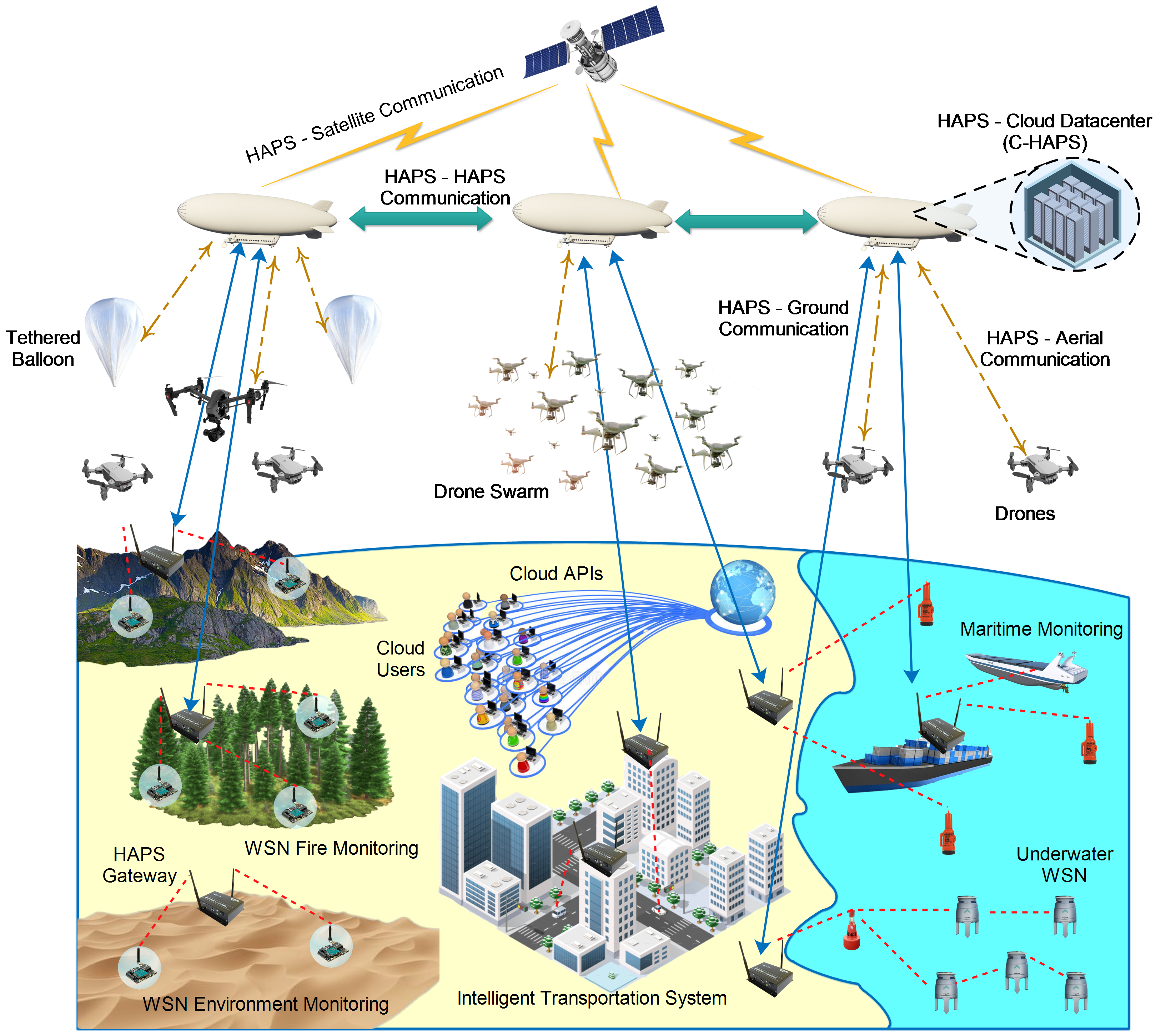}
\caption{\color{black} Conceptual framework for C-HAPS illustrating its components, applications, and communications.}
\label{fig_cHAPS_arch}
\end{figure*}

\section{Background}
\label{Sec_back}

The paper first provides a brief background related to the general prospects of C-HAPS and blockchain technology, before delving into their integration specifics in Section \ref{Sec_proposed}.

\subsection{C-HAPS}
{\color{black}{As highlighted above, the appealing features of HAPS offer a chance for service providers to exploit the HAPS strategic location in reaching out to a broader range of consumers, by means of augmenting HAPS with cloud computing functionalities. }}The system-level prospective of such C-HAPS would further allow the cloud providers to jointly enhance the quality of their services and expand their infrastructure capacities. Given the HAPS long-term vision to connect users in rural and isolated areas (\color{black}see Fig. \ref{fig_cHAPS_arch}\color{black}), C-HAPS has the valuable additional potential to ultra-connect the users in urban areas by offering extra space and resources to increase the users' quality-of-service (QoS). C-HAPS prospects can be further enhanced to provide several applications and services with a multitude of socioeconomic, environmental, and technological impacts. In \cite{mershad2021cloud}, we identified several cloud services that can be efficiently deployed within C-HAPS. Examples of these services include: 


\begin{itemize}
\item 
\textbf{Satellite as a Service:} {\color{black}{Satellite applications can be offloaded to the HAPS cloud. In such cases, customers will be able to consume these services with a better QoS such as reduced delay and enhanced accuracy (such as in localization). }}

\item 
\textbf{Sensor as a Service:} The HAPS network can enhance and enrich current IoT cloud platforms. HAPS ground gateways can connect Wireless Sensor Networks (WSN) to HAPS stations, which is extremely useful to WSNs that are deployed in isolated areas (such as mountains, deserts, oceans, etc.). In addition, tethered balloons (TBs) can increase the coverage area of the HAPS station to cover the WSN network. The system proposed in this paper focuses on enhancing the performance and security of one of the important applications of this service. 
\item 
\textbf{Transportation as a Service:} {\color{black}{HAPS nodes possess high storage and processing capabilities and are suitable for storing and analyzing continuous data streams of smart city applications. In such scenario, HAPS gateways can be placed on high locations (such as roofs of buildings) and connected to roadside units (RSUs). This allows the application to offload ITS data to the C-HAPS and process it fairly quickly. }}

\item 
\textbf{Aerial Network as a Service:} {\color{black}{UAV networks in isolated areas can exploit the existence of a HAPS station to connect to the cloud. Some UAV nodes could be equipped with HAPS gateways to enable this connection. }}

\item 
\textbf{Other services:} Numerous cloud services can be deployed in the C-HAPS once the latter is mature. Examples include Routing as a Service, Gaming as a Service, Social Network as a Service, Crowdsourcing as a Service, etc.
\end{itemize}

\color{black}

To best assess the gains of C-HAPS vis-\`a-vis the ever-increasing security attacks facing the information and communication technology advancements, this paper designs a blockchain system to safeguard future C-HAPS platforms. The paper particularly focuses on the application of EIM using IoT, the details of which are explained in Section \ref{Sec_proposed}.

\subsection{Blockchain}

\color{black}A Blockchain is a special type of distributed database that is built as a series of blocks. 
The blockchain is an append-only database in which blocks are linked via a cryptographic hash. 
Each transaction is signed by its creator and the signature is appended to the transaction for validation. A block consists of a header and body. The header comprises metadata such as the block height, timestamp, hash of the previous block, and Merkle tree root; 
while the body contains the block transactions. 
Blockchain nodes that create the blocks are called miners. 

\color{black}Among the main components of the blockchain is the consensus protocol that allows the blockchain nodes to reach an agreement on various blockchain-related decisions. Mainly, when a new block is generated, the consensus protocol is executed by the blockchain nodes. After the protocol terminates, a single decision should be reached at all nodes, which is usually to add the new block to the blockchain or reject it \cite{mershad2021proof}. A large number of consensus mechanisms have been proposed, such as Proof of Work (PoW), Proof of Stake (PoS), Practical Byzantine Fault Tolerance (PBFT), etc \cite{xiao2020survey}. Each blockchain application requires a specific consensus protocol that suits the applications' needs. In Section \ref{Sec_block}, we describe the consensus algorithm that we propose for the EIM application to guarantee its latency requirements.
 
\color{black}
A large number of sectors and applications have exploited the blockchain to secure their data. Examples of these sectors include finance, healthcare, agriculture, supply chain, etc. Among these sectors is the cloud, as the use of blockchains in cloud computing is one of the growing research directions and is expected to depict a fast expansion across the fields. There are many benefits of blockchain technology concerning cloud computing, including those associated with business data handling, encryption, and privacy. While previous works proposed several methods for implementing the blockchain in ground-based cloud systems; to the best of our knowledge, this paper is the first attempt to integrate the blockchain into C-HAPS. Such integration is expected to strengthen the C-HAPS system by making the data stored in the C-HAPS data center immutable, which would increase the customers' trust in the system. In addition, we propose a permissioned private blockchain network, in which only customers who are certified to access the blockchain service would be able to execute the smart contract of the service and consume it. Moreover, we describe a method that encrypts the sensors' data while in transit. The combination of security measures described in this paper, including the blockchain, secures the C-HAPS system and data to a high degree from major cyberattacks and provides the required characteristics of C-HAPS services, such as authentication, confidentiality, and privacy. 

However, several challenges need to be considered and tackled in order to realize a successful blockchain integration into C-HAPS. Most of these challenges are related to the HAPS characteristics, such as the line of sight (LOS) requirement and the physical maintenance of the HAPS stations. In addition, integrating the blockchain would increase the energy requirements of the HAPS stations. Furthermore, with the expected limited number of HAPS stations that are deployed in C-HPAS, the scalability of the system should be studied as more ground nodes join the blockchain and the number of transactions increases. In this paper, we focus on studying the last aspect and test several scenarios in which the number of cloud users is varied. As illustrated later in Section \ref{Sec_performance}, the proposed system is scalable up to a high degree and its performance is slightly affected as more sensor nodes are added to the network. 

\color{black}

\section{Literature Review}
\label{Sec_review}

In this section we review the state of the art related to the EIM application, focusing on previous research works that proposed blockchain and/or HAPS-based systems for EIM. Dong et al. \cite{dong2015energy} propose an integrated HAP-satellite (IHS) architecture for emergency scenarios with the aim of providing information transfer services for remote sensor devices. In the proposed system, the transmission power requirements of the terminal end and HAP end are investigated in a slow flat Rician fading channel. In addition, an energy-efficient transmission strategy for the energy-efficient path selection is designed using the concept of link-state advertisement (LSA) to reduce energy consumption at both ends.

The authors in \cite{moussa2022earp} propose an Ant Colony Optimization (ACO)-based system for WSNs that are used to detect forest fires. The proposed system considers a multi-sink-based clustered WSN model in which mobile sink(s) can traverse the sensing field to collect data from cluster heads. The main objective is to overcome situations when there is a failure at one or more data collection points and hence achieve high fault tolerance. The ACO algorithm seeks to find the optimal assignment of cluster heads to mobile sinks in order to reduce energy consumption. 

Several papers investigated the idea of utilizing the Internet of Drones (IoD) within the context of the EIM application. Liao et al. \cite{liao2021securing} propose a consortium blockchain system to guarantee trusted collaboration between controllers of software-defined IoD. The system makes use of multi-UAV networks to monitor the environment of a smart city. The IoD cloud services are offered via smart contracts that are managed by a group of drone controllers. In addition, the authors propose the proof of service guarantee (PoSG) consensus protocol in which each service provider elects some trusted controllers as representatives to participate in the consensus process. Next, the controllers elect the block creator and verifiers based on resource guarantees.

A sparsity-optimized spatiotemporal data aggregation model for large-scale disaster monitoring through WSNs is proposed in \cite{li2021blockchain}. The system includes a UAV identity authentication mechanism that ensures the security of transmitted data within the context of a disaster semantic blockchain (DSB). The disaster semantics are obtained by exploring the semantic association relationship of disaster, background, event, and sensor data. Sensor nodes are divided into cluster members, cluster heads, and relay nodes. In addition, data flows are categorized into three levels: multi-cluster, isolated-cluster, and isolated node. A spatiotemporal data aggregation mechanism is performed at the cluster head level, and a temporal data aggregation process is performed by the isolated nodes.  

A blockchain system for distributed Network Function Virtualization (NFV) is proposed in \cite{fu2019resource}. The proposed system (MOMEC) is implemented within the context of Mobile Edge Cloud (MEC) to reach consensus among multiple NFV Management and Orchestration (MANO) systems. The authors formulate the resource allocation problem as a multi-objective optimization challenge by considering the trust features of blockchain nodes and NFV MANO systems, as well as the blockchain computational capability. The authors propose a Deep Reinforcement Learning model to solve the resource allocation problem by improving the blockchain throughput and reducing the cost of providing user services.

The system proposed in \cite{hrovatin2022privacy} utilizes a blockchain model within the context of infrastructure monitoring to collect construction pollutant data via WSN sensors and execute smart contracts to automatically monitor the level of construction pollutants and evaluate the environmental performance. In the proposed system, site inspection forms are uploaded to the blockchain system from the supervisor node and automatically translated to smart contracts. Periodically, the smart contract results are evaluated by the construction peers via the Kafka consensus protocol and added to the blockchain. 

A permissioned blockchain is utilized by the authors in \cite{zhong2022blockchain} to preserve the privacy of data mining WSN nodes. Smart contracts are used to create an onion-like structure comprising the Hoeffding trees and a route. The onion-routed query hides the identity of the sensors from external adversaries, and obscures the data mining results to conceal them from compromised nodes. Hence, a trusted node shares a partially constructed model so that each other sensor node has access to a partial model that is the result of the computation of the previous sensor nodes.

While the discussed systems presented several mechanisms for enhancing the performance of the EIM application via blockchain or HAPS, none of these systems considered a joint HAPS/blockchain model that utilizes the resources of the various network participants to reduce the latency and improve the security of EIM. In the next section, we describe our proposed blockchain-based C-HAPS framework for enhancing the performance and security of the EIM application.

\section{C-HAPS Blockchain System}
\label{Sec_proposed}

\subsection{Problem Statement}
This section presents the proposed C-HAPS blockchain system, specifically designed for EIM applications that utilize the Internet of Things and wireless sensor networks. In such applications, a large number of sensor nodes (SNs) are used to monitor the environment and continuously send their readings to the cloud server via a sink/gateway. Many environment and infrastructure monitoring applications use the above sensing strategy to obtain the SN readings, analyze them, and produce results that are consumed by users via the cloud service API. An example of such applications is fire monitoring, whereby SNs are distributed over the area of a large forest to detect fires at an initial stage. Another example is smart grid infrastructure monitoring in which SNs are deployed at critical points in the smart grid network to monitor for incidents that affect the power system. A third application is related to natural disaster recovery, in which SNs are deployed into drones that fly over the affected area to search for survivors and provide aid whenever needed. 

All applications highlighted above utilize sinks/gateways to send sensors' readings to the cloud. For example, in fire monitoring and smart grid applications, several sinks are deployed at specific locations within the WSN area. In disaster recovery, ground vehicles (which could be fixed or mobile) are used to play the role of the ground control stations (GCS) that connect the drones to the cloud network \cite{mershad2022proact}. From a system level perspective, all such EIM applications are particularly delay-sensitive, since sensors’ readings must be obtained by the user within a certain time limit for further analysis; otherwise, the readings would become obsolete. For instance, readings of temperature and smoke sensors in fire detection applications must be read by the emergency control software within a limited time frame, which allows to respond in a timely fashion to fire events. In smart grid applications, power incidents can be handled by the response tool by isolating the affected areas, the efficacy of which depends on how early the incident is detected. Similarly, in disaster response and recovery applications, a drone that detects a survivor in a critical condition must send its readings to the cloud as fast as possible to enable medical aid and ambulance drones to reach the location of the survivor in a quick manner and provide medical assistance. 
The above emphasizes the fact that EIM applications must meet predefined latency requirements, all while satisfying quality-of-service constraints at the users' sides, defined as a function of data rate and data reliability.

When deploying the EIM application as a cloud service, careful consideration should be made to secure its various elements and aspects. In general, the heterogeneity in the networks and devices that connect to the cloud and the openness of the cloud to various technologies create a challenge related to its security. Recently, a large number of researchers proposed the blockchain as a security solution for cloud services \cite{gai2020blockchain, nguyen2020integration}. Also, the Blockchain as a Service \cite{jie2021research} emerged as one of the best solutions for companies to secure their data. In this paper, we discuss a blockchain framework that is tailored to the C-HAPS EIM application. While integrating the blockchain into the EIM application can provide several benefits in terms of data confidentiality and integrity, it includes many challenges. For example, the blockchain comprises several heavy operations, such as transaction signing and validation, consensus protocol, and smart contracts that make the blockchain a complex system that adds a lot of overhead to the participating nodes. 
Mainly, the processing capabilities of IoT nodes hinder them from engaging in heavy blockchain consensus protocols such as Proof of Work (PoW). In addition, the energy supply of these devices is usually limited. Therefore, they cannot afford to spend a lot of their power executing blockchain programs. Hence, it is essential to make sure that the EIM requirements (such as delay, processing, power consumption, etc.) are met when utilizing the blockchain in these applications. 

\begin{figure*}[!t]
\centering
\includegraphics[width=5.5in]{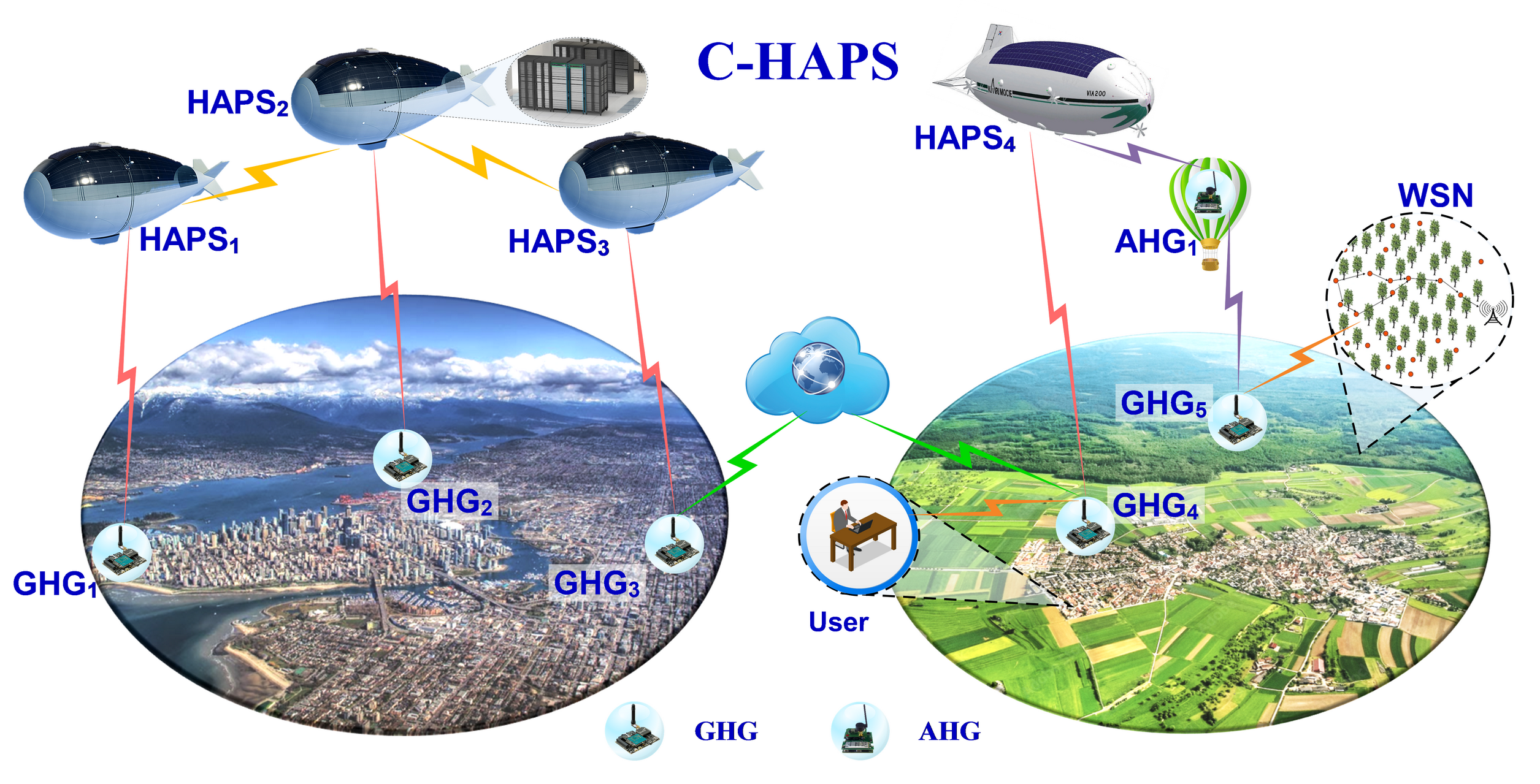}
\caption{The deployment of C-HAPS in two areas: urban (left) and rural (right).}
\label{fig_arch}
\end{figure*}

\subsection{System Architecture}
\label{Subsec_sysarch}

In the proposed model, C-HAPS services are hosted within the HAPS stations datacenter. The HAPS network contains standalone stations that do not connect directly to other stations, in addition to mini-HAPS constellations that contain few HAPS stations. Such mini constellations are usually deployed in large urban areas. In our simulations (Section \ref{Sec_performance}), we show that several HAPS stations are required when providing cloud services to large urban areas. Our model was designed and implemented while considering the delay-sensitive requirement of EIM and the limited resources of sensor nodes. 
The proposed model aims to secure the application data from malicious access and modification via the blockchain, while performing the block generation and consensus in a fast manner that satisfies the delay requirements and poses negligible overhead on the SNs. In addition, we show that our proposed model is resilient to general attacks on cloud applications and blockchain-specific attacks. First, We start by describing the network architecture. 
The nodes that participate in the proposed system are: 
\begin{itemize} 
\item 
\textbf{Sensor nodes:} 
\color{black}Based on the application, sensor nodes could be deployed in urban areas (for example, within homes, vehicles, companies, factories, grid infrastructure, etc.) or rural ones (for instance, forests, mountains, oceans, deserts, etc.). In this paper, we consider SNs that are used within the EIM context. EIM applications usually require a large number of SNs that are deployed over a wide area, creating several WSNs. \color{black} A sensor node senses its environment, generates data, and sends it to the sink at a predefined frequency either directly or via other SNs. A large number of routing protocols for WSN and IoT networks were proposed in the literature, with the aim of delivering the data from the sensor node to the sink in an efficient manner that satisfies the objectives of the application (for example, to achieve small delay, low energy consumption, high reliability, etc.) \cite{bhushan2019routing, mershad2020surfer, mershad2020blockchain}. The frequency at which a sensor node sends its data depends on the application requirements and can be dynamically adjusted by the network administrator in most IoT/WSN applications \cite{anastasi2009extending}.  
\item 
\textbf{Ground-HAPS gateway:} \color{black}A WSN usually contains special nodes that are connected to the application backend. These nodes, called sinks, act as middleware between the SNs and cloud servers. Depending on the WSN size and objectives, sinks are usually deployed at different locations within the WSN to ensure that all SNs can connect to the backend. \color{black}In our model, we integrate a HAPS communication module into the sink such that it forwards the readings that it receives to the HAPS station instead of the ground cloud server. Hence, the sink becomes a ground-HAPS gateway (GHG) that bridges the connection between the HAPS and any ground node that is not equipped with a HAPS communication module. Also, the GHG provides a means for a standalone HAPS station to connect to the ground cloud and to other HAPS stations. In general, the GHG is equipped with better resources than sensor nodes (processing, storage, and energy supply). Hence, it can participate in some of the blockchain operations, as we discuss in the next section.  
\item 
\textbf{Aerial-HAPS gateway:} Many applications that utilize aerial vehicles such as drones contain specific drones that act as aerial gateways that connect the ground control station (GCS) to the drones that are outside the communication range of the GCS. In some types of applications, drones may fly to locations that are outside their movement path based on the incidents that occur during their missions. Some other applications do not define a movement path for drones. In all these cases, it would be necessary to deploy one or more drones as aerial gateways that fly in such a way to maintain the connection between these drones and the GCS. \color{black}Hence, aerial gateways are usually deployed when the aerial network contains nodes that do not have a connection to the GCS. In such cases, one or more aerial gateways are used to play the role of an aerial sink.   
\color{black}In our system, we equip these aerial gateways with HAPS communication modules such that they can send drones' readings to the HAPS stations. In other scenarios, tethered hot air balloons can be equipped with HAPS communication modules as well. These aerial nodes become aerial-HAPS gateways (AHG) that bridge the connection between ground and aerial nodes and the HAPS network. The AHG can also be used to connect a standalone HAPS station to other HAPS stations when needed. Fig. \ref{fig_arch} shows the deployment of C-HAPS stations in urban and rural areas and the utilization of GHGs and AHGs. In the urban area, three HAPS stations HAPS\textsubscript{1}, HAPS\textsubscript{2}, and HAPS\textsubscript{3} are positioned over a city and its suburbs. GHG\textsubscript{1} connects to HAPS\textsubscript{1}, GHG\textsubscript{2} to HAPS\textsubscript{2}, and GHG\textsubscript{3} to HAPS\textsubscript{3}. In the rural area, a single HAPS station (HAPS\textsubscript{4}) is installed. GHG\textsubscript{4} connects directly to HAPS\textsubscript{4}, while GHG\textsubscript{5} is outside the range of HAPS\textsubscript{4} and connects to it via AHG\textsubscript{1}. HAPS\textsubscript{4} is a standalone station and can connect to the other HAPS stations via the gateways. 
\item 
\textbf{HAPS station:} This node communicates with GHGs and AHGs to collect data from SNs, save them into the blockchain, and provide them to cloud users. The HAPS station plays the role of a full blockchain node that stores the full blockchain, participates in generating new blocks and validating other’s blocks, and responds to users’ requests by executing blockchain smart contract functions and sending blockchain data to users in a secure manner.
\item 
\textbf{Cloud user:} A cloud user accesses the cloud application at the HAPS via his/her cloud account. The user authenticates him/herself via the blockchain smart contract and obtains a certificate that defines the user’s roles and privileges. The user provides the certificate alongside each blockchain request that is sent to the HAPS station. Based on the certificate, the user can retrieve IoT data from the blockchain, execute specific smart contract functions to obtain results from the data, and/or adds his/her own transaction to the blockchain via the consensus model which is explained shortly. 
\end{itemize}

\begin{figure*}[!t]
\centering
\includegraphics[width=5.5in]{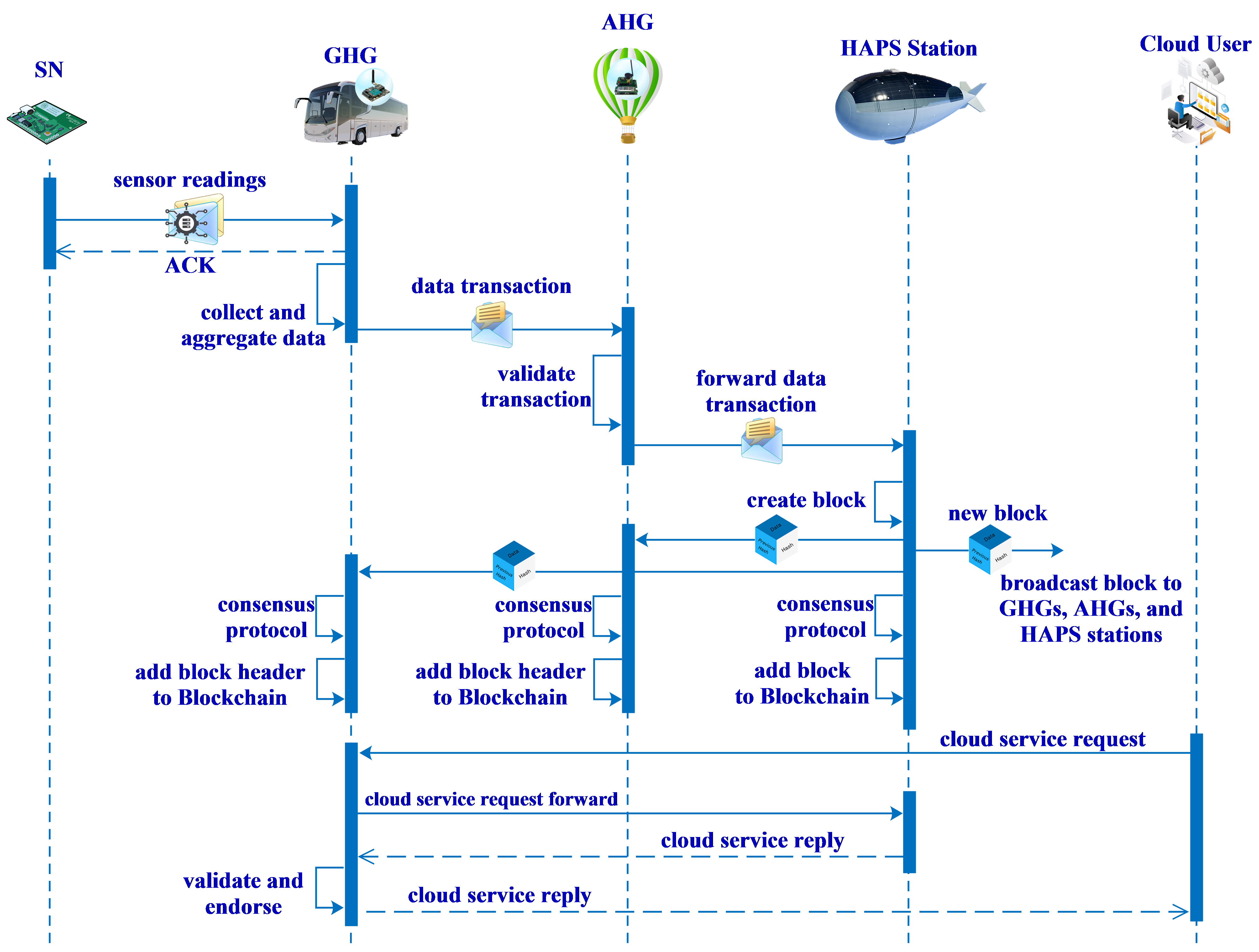}
\caption{\color{black}General communications between the various system nodes.\color{black}}
\label{fig_uml1}
\end{figure*}

\color{black}
Figure \ref{fig_uml1} illustrates the general communications between the system nodes. As shown in the figure, sensor nodes send their readings to the nearest GHG/AHG via the WSN. Each GHG/AHG validates the data it receives from SNs, aggregates them, and sends data transactions to the HAPS stations, possibly via other GHGs/AHGs. The HAPS station receives data transactions from GHGs/AHGs and saves them in its pending list. When its turn arrives, the HAPS station creates a new blockchain block and broadcasts it to the network of GHGs/AHGs and HAPS stations. Each node among the latter executes the consensus protocol (more details about it later) and updates its copy of the blockchain. As we explain later, the HAPS stations store the whole block while GHGs/AHGs store only the block header. When a cloud user wants to consume the cloud service, it sends a cloud service request to the nearest GHG/AHG, and the latter forwards the request to the nearest HAPS station that hosts the required service. The HAPS station replies to the GHG/AHG which validates the reply based on its local copy of the blockchain, endorses the reply packet, and sends it to the user.  

\color{black}
Before we discuss the details of the proposed system, we describe in the next section the adversary model that we consider in our system.

\subsection{Adversary Model}
\label{Sec_adversary}

Among the various types of nodes that were described in the previous section, we consider the HAPS station to be fully trusted. We assume that the cloud services that are deployed and offered from the HAPS station datacenter are highly secured against various attacks. 
On the other hand, we assume that the GHG/AHG is semi-trusted. While these devices are usually secured as much as possible, attackers could still be able to exploit software vulnerabilities or physically break the device (in certain situations) to compromise it and use it to launch attacks. Hence, we describe an approach that is applied by the proposed consensus protocol to detect and defend against a compromised gateway. Finally, we assume that sensor nodes are the least trustworthy nodes. Due to the limited resources of these devices and their existence in deserted locations, they are more susceptible to cyberattacks. Hence, a mechanism should be applied by the sink to detect the false data that would be transmitted by a compromised sensor node. A large number of papers proposed various solutions to this problem \cite{xie2011anomaly, alotaibi2019security, anand2022detection, wang2021active}. In this paper, we aim at presenting solutions to the attacks that could occur with the implementation of the blockchain in the Cloud-HAPS network.  Mainly, we focus on the following attacks:
\begin{itemize}
\item 
\textbf{Transaction Piracy attack:} this type of attack occurs when the sensor node sends secret or private data to the blockchain and an attacker is able to steal the data.  
\item  
\textbf{False Transaction attack:} in this type of attack a malicious sensor node or gateway attempts to inject a false transaction into the blockchain by either creating a fake transaction or tampering with a valid transaction that it receives from another node. 
\item 
\textbf{Block Withholding attack:} this attack occurs when the blockchain miner that is generating the new block executes a malicious action to stop or delay sending the new block to the consensus nodes with the aim of delaying the consensus process and affecting the application performance. 
\item 
\textbf{51\% or Majority attack:} this attack occurs when the attacker is able to compromise a large number of blockchain miners with the aim of overriding the consensus algorithm and breaking the blockchain integrity by inserting false blocks or transactions into the blockchain without being detected.  
\item 
\textbf{Forking and Replay attack:} a blockchain fork occurs when the consensus process generates two or more different blocks that are selected by the different consensus nodes. Hence, multiple blockchain branches appear. Attackers exploit this incident by executing the same malicious transaction in multiple branches to gain benefits or affect the blockchain integrity.   
\item 
\textbf{Consensus Sabotage attack:} this attack is performed by a malicious blockchain node that participates in the blockchain consensus process and attempts to sabotage it by performing various malicious actions.
\item 
\textbf{Consensus Collusion attack:} this attack is performed by a group of nodes that participate in the consensus process and collude to perform malicious actions with the aim of destroying the consensus process or compromising the blockchain integrity. 
\end{itemize}

\subsection{Transaction Generation and Validation}
\label{Sec_trans}
When a sensor node generates new readings, it combines them into a data packet, signs it, and sends it to the nearest sink. If the data is private, the sender encrypts the data with the receiver's private key. The data packet passes through zero or more sensor nodes on its way (based on the sender’s location and the routing protocol). Each sensor node on the packet’s path validates the packet by checking the correctness of the data (for public data only) and the validity of the sender’s signature. If the packet is valid, the intermediate sensor node endorses the packet by adding its signature to the packet before forwarding it to the next sensor node or to the sink. 

\begin{figure*}[!t]
\centering
\includegraphics[width=5.0in]{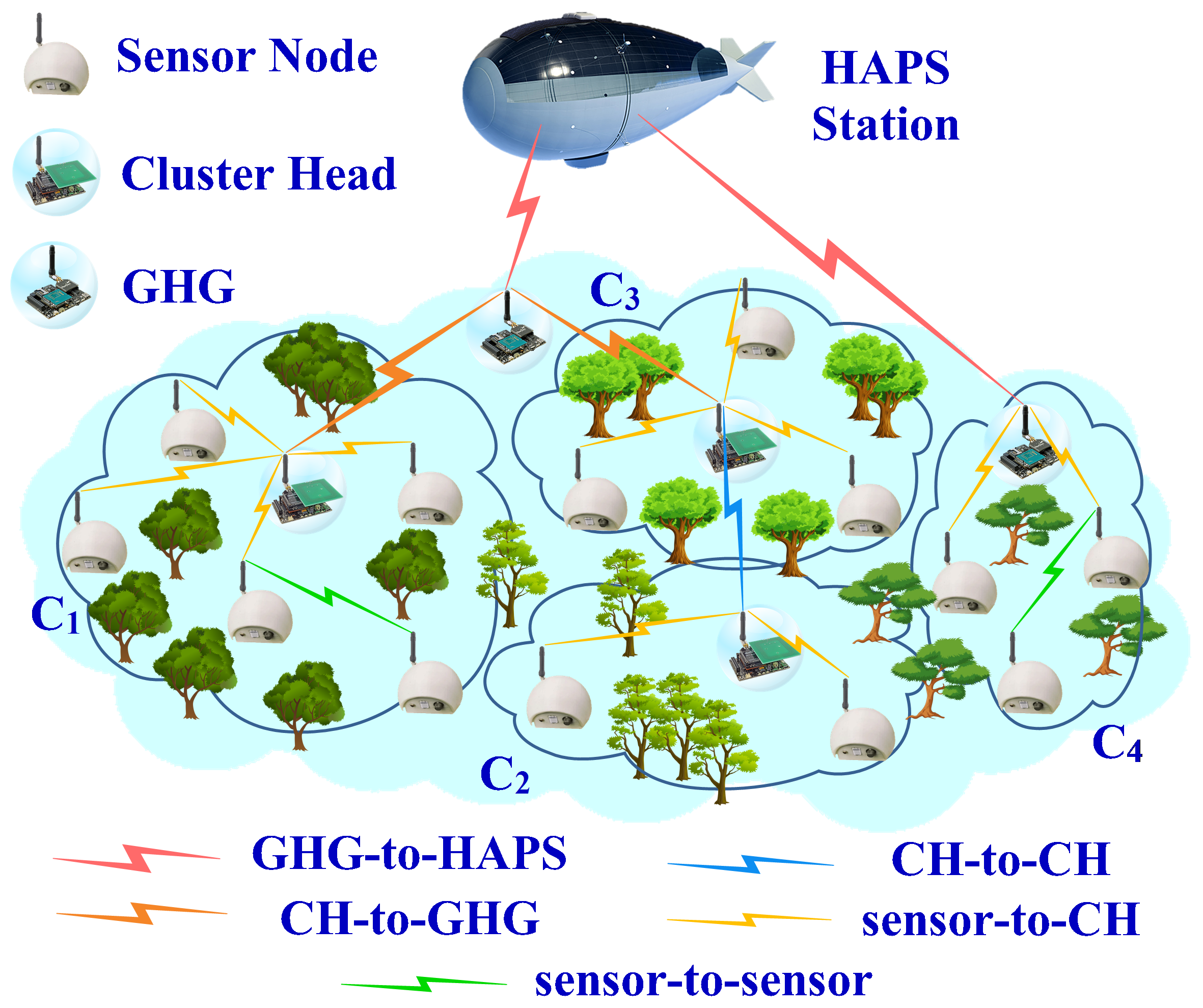}
\caption{The architecture of the WSN in the proposed application.}
\label{fig_wsn}
\end{figure*}

Note that sensor nodes are usually clustered based on their geographic locations such that each node maintains the information (such as the ID, location, and public key) of all nodes in the cluster. Also, in such applications, one of the nodes in the cluster is selected as the cluster head (CH) that is responsible for forwarding all packets from/into the cluster. In some types of sensor networks, each CH is a sink. In other applications, each group of cluster heads connect directly to a sink. Several clustering approaches were proposed in the literature \cite{shahidinejad2020sink, amini2019improving, ahmed2018cbe2r}. Our proposed framework is general and can be used with any clustering mechanism. In the simulations (Section \ref{Sec_performance}), we clustered the SNs based on the approach proposed in \cite{amini2019improving}. The architecture of the proposed sensor network is illustrated in Fig. \ref{fig_wsn}. In the figure, sensor nodes are divided into four clusters C\textsubscript{1}, C\textsubscript{2}, C\textsubscript{3}, and C\textsubscript{4} with a cluster head in each cluster. Sensor nodes that are near to the CH send their readings directly to it, while sensor nodes that cannot connect directly to the CH send their reading via other sensor nodes (as in C\textsubscript{1} and C\textsubscript{4}). Some CHs, such as that of C\textsubscript{4}, could be equipped with a HAPS communication module and hence act as a GHG. Other CHs send their aggregated data to the nearest GHG (C\textsubscript{1}, C\textsubscript{2}, and C\textsubscript{3}). Some CHs connect directly to the GHG, while other CHs (such as that of C\textsubscript{2}) connect to the GHG via other cluster heads.

When a GHG receives a packet from a sensor node or a CH, it checks the validity of the data by verifying the endorsement signatures that were added by the intermediate nodes and the signature of the sender. Based on the data that the sink received from other sensor nodes, the sink could aggregate the received data with the existing data, delete the received data (if it is a duplicate of an existing data record), or save the data for possible future aggregation. Each GHG periodically (based on the application requirements, for example, every 50 milliseconds) aggregates and groups the existing data into a blockchain transaction packet, signs it using its digital signature, and sends it to the nearest HAPS station that hosts this cloud service. In some cases, the packet is sent directly from the GHG to the HAPS station. In other cases, the packet could pass through one or more intermediate GHSs/AHGs before reaching the HAPS station that hosts the application.  

Note that in the proposed system, each node encrypts a transaction packet with the next node’s public key before sending it. For example, consider a scenario in which GHG G\textsubscript{1} sends a transaction packet to the HAPS station H\textsubscript{1}. The packet is sent on the following path: {G\textsubscript{1}, G\textsubscript{2}, A\textsubscript{1}, H\textsubscript{1}}. Here, G\textsubscript{1} and G\textsubscript{2} are GHGs and A\textsubscript{1} is an AHG. Hence, G\textsubscript{1} encrypts the packet with G\textsubscript{2}’s public key. G\textsubscript{2} receives the packet, decrypts it, validates G\textsubscript{1}’s signature, encrypts the packet with A\textsubscript{1}’s public key, and sends the packet to A\textsubscript{1}. In its turn, A\textsubscript{1} decrypts the packet using its private key, validates G\textsubscript{1}’s signature, encrypts the packet with H\textsubscript{1}’s public key, and sends the packet to H\textsubscript{1}. Also, note that in this paper we do not focus on the routing protocol that is used to route packets between GHGs/AHGs and the HAPS network and vice versa, as we plan to describe its details in a future paper. We assume that the routing protocol is responsible for routing packets between nodes when there is no direct link between them. For example, when the CH does not have a direct link to a GHG, the routing protocol is used to calculate the nearest GHG and find the path to it.

Each GHG and AHG stores a pending transaction list in which it saves all transactions that it generates/forwards before they are added to the blockchain. A transaction remains in the GHG/AHG pending list until the latter receives the next blockchain block. At this stage, the GHG/AHG checks its pending list and removes from it the transactions that have been successfully added to the blockchain. Also, each transaction in the pending list has an expiry deadline at which it would be deleted if it is still not added to the blockchain. The expiry deadline defines the time at which the transaction becomes obsolete and should be discarded.  

In the proposed system, the GHGs, AHGs, and HAPS stations form the blockchain network. However, only HAPS stations are responsible for generating new blockchain blocks, while all the blockchain nodes (i.e., GHGs, AHGs, and HAPS stations) participate in the consensus protocol to validate the new blocks. This distinction was made based on the fact that the HAPS stations are the final destinations of all data transactions. On the other hand, the transactions will be scattered between the GHGs and AHGs. Hence, the HAPS station can group the transactions that it receives from the gateways, generate the new block by ordering the transactions based on their timestamps, generate the Merkle root and other parameters of the block header, and add the hash of the previous block. In order to satisfy the delay constraints of the EIM application, this operation is performed frequently such that transactions are added to the blockchain within the delay threshold of the application. More details about this issue will be discussed in the next section. With respect to the GHGs/AHGs, each gateway participates in the consensus protocol by validating that the transactions in its pending list were added correctly to the new block. 

\subsection{Blockchain Architecture}

Several cloud services can be offered by C-HAPS. Each service will have its own blockchain that is configured based on the service characteristics and requirements. Many cloud services require a private blockchain that can be accessed by the service administrators and subscribers only. Other cloud services could use a hybrid blockchain architecture and restrict access to the private features of the service by utilizing smart contracts. In such cases, the users can access the paid features only after they subscribe to the service and their credentials are added to the blockchain. Hence, the smart contract checks the user’s credentials and identifies his/her access rights before enabling the user to access the private features. In certain cases, applications that share the same characteristics and settings can share a single blockchain and distinguish their transactions via a unique service identifier that is added to each transaction. 

With respect to blockchain data, different types of cloud services store data with different sizes and characteristics. The blockchain administrators usually have the option to store the transactions directly in the blockchain block (on-chain), or to save the data in an external dedicated storage system (such as the InterPlanetary File System or IPFS) and save the transactions’ hash and metadata in the blockchain (off-chain). This latter option is more efficient when the service uses or generates enormous amounts of data. In such cases, the external storage system can provide better indexing and searching capabilities than the blockchain. Hence, the data can be retrieved faster from the external storage system and validated via the hashes in the blockchain. Several recent works proposed a variety of models that are based on off-chain storage \cite{xu2021slimchain, jayabalan2022scalable}. 

\subsection{Block Generation and Consensus}
\label{Sec_block}

\subsubsection{Block Creation} 

In the proposed blockchain system, blocks are generated by HAPS stations at constant time intervals. Each cloud service defines its block generation time interval based on the application delay requirements. 
The HAPS stations that host a certain EIM service generate the blocks of that service successively based on their network IDs. Each time a HAPS station receives a transaction, it validates it by checking the signatures of the sender and endorsers. Next, it broadcasts the transaction to all the HAPS stations that host this service. Each HAPS station validates the transaction and saves it in its pending transaction list.

When the block interval timer expires, 
the HAPS station whose turn to generate the new block orders the transactions in its pending list based on their timestamps, creates the new blockchain block, and broadcasts it to all the HAPS stations that are hosting this EIM service in addition to all GHGs and AHGs who are forwarding the transactions packets of the EIM service. Note that each HAPS station saves a list that contains the IDs of all GHGs/AHGs that are participating in the blockchain of each EIM service that is hosted by the HAPS station. The latter updates the list when a GHG/AHG joins or leaves the blockchain network of that service. 

\subsubsection{QUICO Consensus Protocol} 

\color{black}In the proposed blockchain model, the consensus protocol is executed among the HAPS stations that host the corresponding cloud service and the GHGs/AHGs that create and forward the data transactions of this service. The HAPS stations play the role of full blockchain nodes, while the gateways are light nodes that perform part of the consensus operations only, as we will explain in this section. On the other hand, SNs do not participate in the consensus process in order to preserve their limited resources. 
\color{black}

When a blockchain node \textit{N} (which could be a HAPS station or a GHG/AHG) receives a packet that contains a new block, it executes the C-HAPS consensus protocol, which we call QUICO (an abbreviation for quick consensus). The details of QUICO are: when \textit{N} receives a new block from a HAPS station \textit{H}, it verifies that the transactions in its pending list that exist in the new block were added correctly by \textit{H}. If this is the case, \textit{N} sends a “Block ACK” message to \textit{H} that contains the block sequence number and signature of \textit{N}. However, if \textit{N} discovers that one or more transactions in its pending list were added incorrectly to the new block 
(for example, if an error exists in the transaction data or metadata as compared to the transaction in \textit{N}'s pending list), 
then \textit{N} sends a “Block ERROR” message to \textit{H}.

After broadcasting the new block to the blockchain nodes, \textit{H} waits for a certain time to receive the replies of the blockchain nodes. \textit{H} considers the block validated by the blockchain network after it receives a “Block ACK” from all the HAPS stations in the blockchain network and from more than half the gateways. It is important that \textit{H} receives an acknowledgment from each other HAPS station in the blockchain since these nodes hold copies of all pending transactions and should be able to verify all the transactions in the new block. In addition, \textit{H} needs to acquire confirmation from the majority of GHGs and AHGs in the blockchain network. Suppose that the blockchain network contains \textit{X} HAPS stations and \textit{Y} GHGs and AHGs. Also, suppose that the “Block ACK” packet from a HAPS station is labeled as HACK and that from a GHG/AHG is labeled as GACK, then \textit{H} should receive (\textit{X}-1) HACKs and \textit{floor}((\textit{Y}/2+1)) GACKs before it labels the new block as confirmed.

When the new block is confirmed, \textit{H} aggregates the signatures in the confirmations that it received into a “Block CONFIRM” packet and broadcasts it to the blockchain nodes. The other HAPS stations verify the signatures and add the new block to the blockchain. On the other hand, the GHGs and AHGs add the header of the new block to their headers blockchain. In our proposed system, GHGs/AHGs act as light blockchain nodes that store only the headers of the blockchain blocks and use them to validate data transactions that they obtain from the HAPS stations. A light blockchain node in our system stores locally the headers of the blocks only. This concept is similar to the Simplified Payment Verification (SPV) node in Bitcoin \cite{le2019lightweight}. 
When a cloud user requests a blockchain transaction from a HAPS station, the transaction is sent by the HAPS to the GHG/AHG network and then to the user (unless the user has a HAPS communication module, then the user can obtain the block directly from the HAPS station and perform the validation locally). 
Each GHG/AHG that forwards a transaction verifies it using its headers blockchain. If the transaction is valid, the GHG/AHG endorses the transaction by adding its signature to the data packet. This strategy increases the user’s trust in the obtained data, since it will be validated by multiple nodes, which confirms its legitimacy. 

\begin{figure*}[!t]
\centering
\includegraphics[width=5.5in]{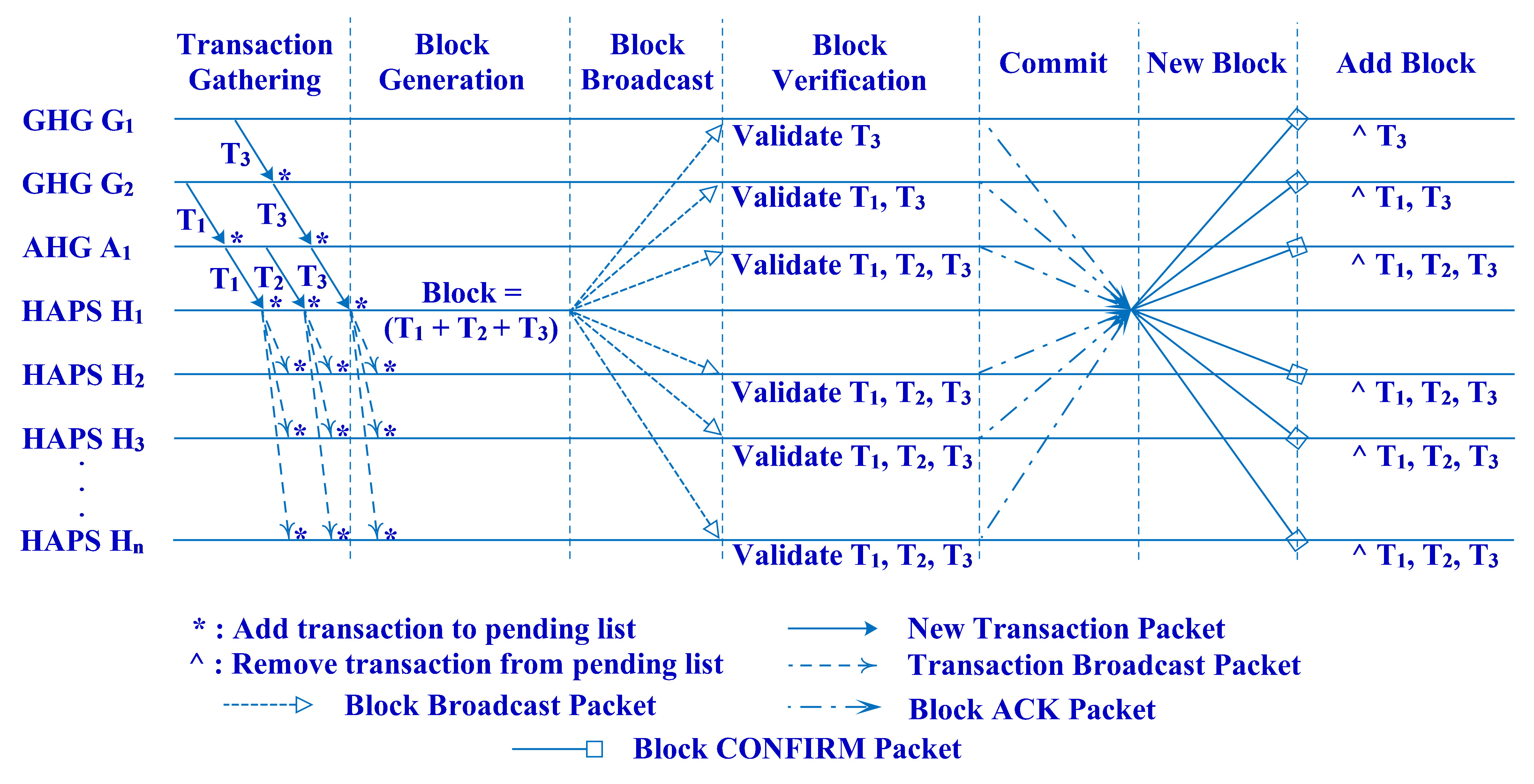}
\caption{A sample scenario of transactions’ gathering, block generation, and consensus.}
\label{fig_consensus}
\end{figure*}

\subsubsection{Sample Consensus Scenario}

A sample scenario is shown in Fig. \ref{fig_consensus}. In the figure, the turn is on HAPS H\textsubscript{1} to generate the new block. Hence, all GHGs/AHGs send their new transactions to H\textsubscript{1}. Suppose that AHG A\textsubscript{1} has a direct link to H\textsubscript{1}, while GHG G\textsubscript{2} sends its packets to H\textsubscript{1} through A\textsubscript{1}, and GHG G\textsubscript{1} sends its packet to H\textsubscript{1} through the path G\textsubscript{2}-A\textsubscript{1}. Three transactions are sent to H\textsubscript{1}: T\textsubscript{1} is sent by G\textsubscript{2}, T\textsubscript{2} by A\textsubscript{1}, and T\textsubscript{3} by G\textsubscript{1}. As shown in the figure, each GHG/AHG that forwards a transaction adds it to its pending list (represented by the symbol “*” in the figure). When H\textsubscript{1} receives a transaction, it broadcasts it to the other HAPS stations. When the time to generate the new block is due, H\textsubscript{1} creates the block and broadcasts it to all GHGs/AHGs and HAPS stations. Each node validates the transactions in its pending list that are in the new block and sends a “Block ACK” packet to H\textsubscript{1}. After H\textsubscript{1} collects the “Block ACK” packets, it broadcasts a “Block CONFIRM” packet to all nodes. Each node that receives a “Block CONFIRM” adds the new block to the blockchain and removes the transactions that are in the new block from its pending list.

\subsubsection{Handling Block Errors}

Going back to QUICO, if the block creator \textit{H} receives a “Block ERROR” from one or more blockchain nodes, it checks the transactions that were labeled as erroneous in the “Block ERROR” packets. If \textit{H} discovers that these transactions were indeed added incorrectly to the block, it regenerates the block by replacing the erroneous transactions with their correct versions and broadcasts the new block (with a new sequence number). 
However, if \textit{H} believes that one or more transactions that were labeled as erroneous by other nodes are correct, it adds the two versions of each transaction to an “ERROR Check” packet and sends the latter to the GHGs/AHGs who endorsed the transaction. Note that a transaction contains the signatures of all GHGs/AHGs who endorsed it. 
A GHG/AHG that receives an “ERROR Check” packet reviews its pending list to see which of the two versions of the questionable transaction matches the transaction in its pending list. The GHG/AHG replies to \textit{H} with an “ERROR Resolve” packet that indicates the correct version with the GHG/AHG signature. \textit{H} collects the GHGs/AHGs replies, resolves the error, and forwards the replies and the decision to the blockchain nodes that sent the “Block ERROR” packets. The blockchain nodes update their pending lists accordingly. After \textit{H} validates all the questionable transactions, it generates a new block 
and broadcasts it to the blockchain nodes, and the process is repeated. 

After the new block is added to the blockchain, the HAPS station whose ID is next in the group of HAPS stations who are hosting this blockchain examines the timestamp \textit{t\textsubscript{b}} in the “Block CONFIRM” packet and starts a timer such that the timer will fire at a time \textit{t\textsubscript{c}} in the future, where \textit{t\textsubscript{c}}-\textit{t\textsubscript{b}} = \textit{t\textsubscript{th}}. Here, \textit{t\textsubscript{th}} is the block generation interval that is defined by the application delay requirements. When the timer expires, the HAPS station performs the same operation that was explained before. 

\begin{figure*}[!t]
\centering
\includegraphics[width=4.8in]{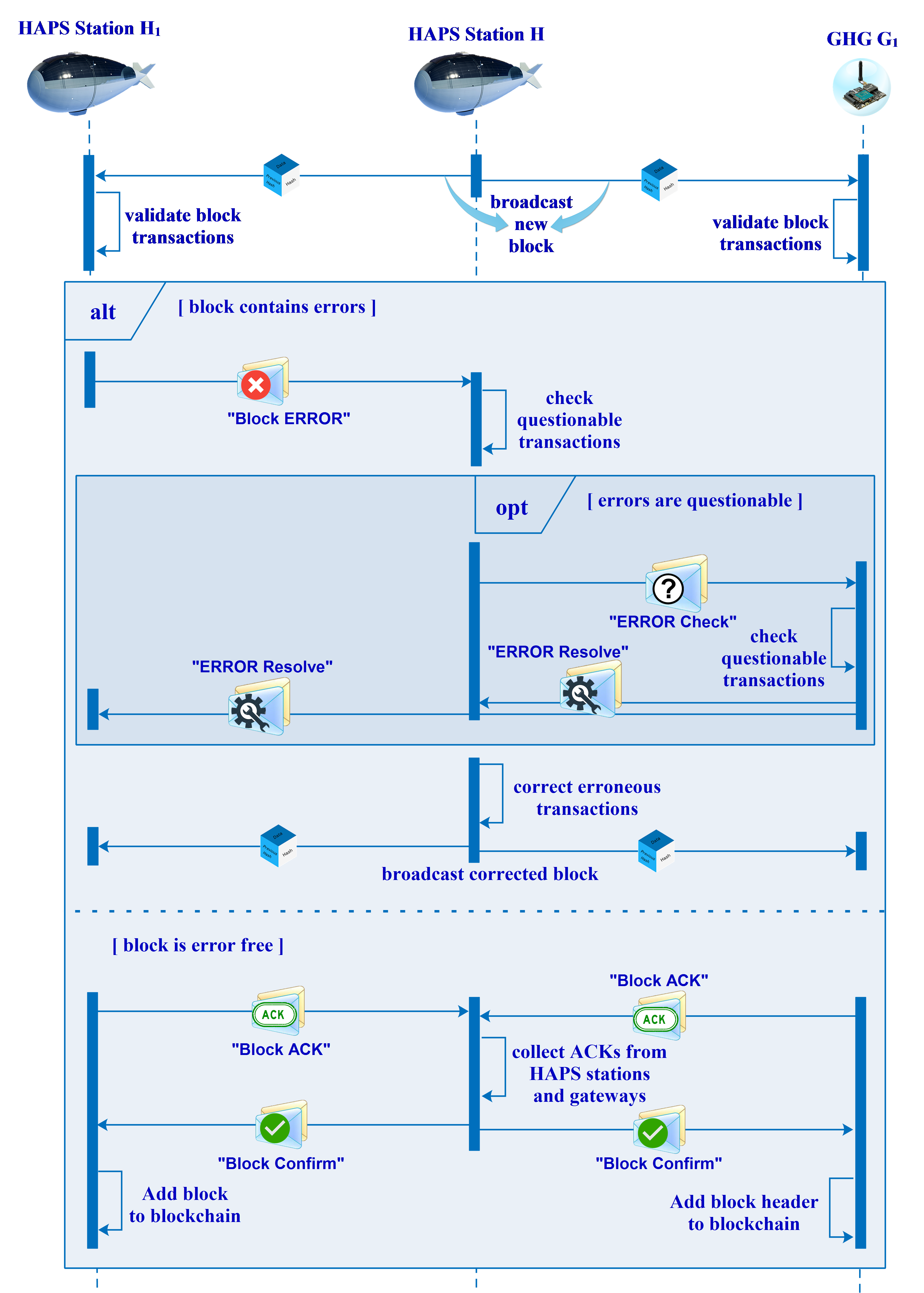}
\caption{\color{black}An illustration of the messages exchanged during the execution of the QUICO consensus protocol.\color{black}}
\label{fig_uml2}
\end{figure*}

Note that the main objective of the QUICO protocol is to generate the block in a time less than the delay requirements of the application. The block creator waits for a small period (\textit{t\textsubscript{w}}) to receive the other’s replies. If the reply of a certain gateway is delayed or dropped, it will not affect the consensus process as long as the number of such replies is less than half the total number of GHGs/AHGs. In normal conditions, the block creator should receive “Block ACK” packets from more than half the GHGs/AHGs within the small waiting time and confirm the new block. In abnormal conditions, such as when an attacker compromises one or more GHGs/AHGs, the block creator could receive “Block ERROR” messages from one or more GHGs/AHGs and execute the second part of QUICO related to validating the questionable transactions, which will delay the consensus process. When the latter case occurs, the block creator sends a consensus report to the cloud service administrators who investigate the reasons that lead to the erroneous transaction(s) at those GHGs/AHGs. This helps the administrators detect the attacker quickly and isolate the GHG/AHG or reset it.

\color{black}
Figure \ref{fig_uml2} illustrates the messages exchanged during the execution of QUICO. In the figure, the HAPS station in the middle (i.e., H) generates a new block and broadcasts it to the blockchain network. H\textsubscript{1} is a HAPS station that receives that new block from H. If H\textsubscript{1} finds no errors in the new block, it sends a “Block ACK” packet to H. If H\textsubscript{1} suspects that one or more transactions in the new block are erroneous, it sends a “Block ERROR” to H. The latter inspects the questionable transactions. If H is not able to confirm whether there is an error or not, it sends a “Block Check” packet to the GHGs/AHGs from which the questionable transactions were generated (represented as GHG G in the figure). Here, G resolves the situation as previously explained and sends an “ERROR Resolve” packet to the HAPS stations. Finally, H regenerates the new block after correcting the erroneous transactions and broadcasts the corrected block to the blockchain network. 
\color{black}

\subsection{Security Analysis}
\label{Sec_Secanal}

In this section, we discuss the security of the proposed blockchain model and its resilience to the attacks that were described in Section \ref{Sec_adversary}. Note that, as we mentioned before, we focused on the attacks that are related to the blockchain. On the other hand, attacks that target the WSN or IoT network have been extensively studied and a large number of solutions already exist (for example, \cite{butun2019security, haque2021security}). 
On the other hand, we focus here on describing how the proposed blockchain system can be used to defend against the following attacks: 
\begin{itemize}
\item 
\textbf{Transaction Piracy attack:} in order to prevent attackers from stealing private data in the sensor nodes’ transactions, encryption is used to protect all such transactions. As described before, each node encrypts each private transaction (including the signature) with the public key of the receiver.
\item 
\textbf{False Transaction attack:} an adversary will not be able to inject a false transaction into the blockchain since it will not be endorsed by the other nodes in the cluster and by the cluster head before it is sent to the sink. In our model, the trustworthiness of the data transaction depends on the endorsements that it receives. If the transaction passes through one or more sensor nodes before reaching the sink, these sensor nodes check the correctness of the transaction data before endorsing the transaction. If the transaction sender is a neighbor of the cluster head or the sink, then the latter performs the same operation. Hence, the only way for an attacker to insert a malicious transaction is by compromising the sender, the nodes between the sender and the sink, and the sink itself. This should prove to be a hard process to be done by the attacker, since the network administrators should detect the attack before the attacker is able to compromise multiple nodes and insert the transaction into the blockchain. 
\item 
\textbf{Block Withholding attack:} based on our assumption that the HAPS station is fully trusted, and since the HAPS stations are the only nodes that act as blockchain miners in our proposed system, this attack should not occur. 
\item 
\textbf{51\% or Majority attack:} this attack occurs if an attacker is able to compromise more than half of the GHGs/AHGs that participate in the blockchain. In our system, if a GHG/AHG is suspected of behaving maliciously, the network administrators remove it from the blockchain network. Hence, as long as the HAPS station is secure, it always reports malicious incidents to the network administrators. When the HAPS station detects any malicious incident by one or more GHGs/AHGs, it delays the block generation process until the issue is resolved by the administrators. This method prevents an adversary from corrupting the blockchain even if it compromises the majority of gateways participating in the blockchain consensus.
\item 
\textbf{Forking and Replay attack:} since only a single block is generated at a time by a single HAPS station, forking cannot happen in the proposed system, which eliminates the possibility of Replay attacks. 
\item 
\textbf{Consensus Sabotage attack:} similar to the 51\% attack, this attack is not possible as long as the HAPS station is secure. If an attacker compromises a GHG/AHG and attempts to sabotage the consensus process, it will be ejected from the blockchain network by the administrators. For example, if the malicious gateway sends a “Block ERROR” packet while there are no erroneous transactions in the new block, the HAPS station reports the incident to the administrators, as described in the previous section. If the GHG/AHG tampers with the new block that is broadcast by the HAPS station, the legitimate gateways will detect the tampered block from the false data in the block body (invalid hash and signatures) and report the incident to the HAPS station. Hence, a malicious gateway will not be able to jeopardize the consensus process as long as the HAPS stations are secure and there exist legitimate gateways in the blockchain network. 
\item 
\textbf{Consensus Collusion attack:} similar to the previous attack, this attack will be detected as long as there exists at least one legitimate gateway that detects the malicious operations of the colluding gateways and reports the incident to the HAPS station. Since all consensus operations are distributed among all the GHGs/AHGs that participate in the blockchain, a group of GHGs/AHGs will not be able to collude and perform an attack on the consensus process without being detected.
\end{itemize}

\section{Performance Evaluation}
\label{Sec_performance}

In this section, we test the proposed system to prove its applicability and efficiency. Currently, there is still no standard testing platform or simulator for HAPS. Hence, we created a software module for the HAPS that we integrated into the network simulator 3 (NS-3) software. This module is loaded into a simulated HAPS node and includes all the functionalities required by a C-HAPS service. The HAPS module includes the parameters for two communication channels: HAPS-to-HAPS (H2H) and HAPS-to-Ground (H2G). These parameters are based on the Electronic Communications Committee Report\footnote{https://docdb.cept.org/download/624}. We also utilized some of the parameters mentioned by the International Telecommunication Union\footnote{https://www.itu.int/rec/R-REC-F.1891/en} and ATIS\footnote{https://www.atis.org/wp-content/uploads/3gpp-documents/Rel16/ATIS.3GPP.38.821.V1600.pdf} to complete the wireless setup of the HAPS module. 

The proposed blockchain system was built using Hyperledger Iroha. We chose Iroha for several reasons: first, it is written in C++, which facilitates the utilization of parts of its source code directly into the NS-3 code. Second, Iroha utilizes the YAC consensus protocol, which is an enhanced version of PBFT that produces lower latency and higher throughput. The proposed QUICO protocol is similar to YAC in terms of the main principles of distributed voting and committing of new blocks. However, QUICO is tailored for the HAPS environment in terms of the specific roles of the HAPS stations and gateways and hence the existence of two different types of votes. In addition, the reject decision in YAC is replaced with the “ERROR Check” and “ERROR Resolve” method in QUICO. Hence, in order to avoid writing the code of QUICO from scratch, we based its implementation on that of YAC while making the necessary adjustments and additions. Overall, we created another NS-3 module that contains the blockchain operations that were discussed in Sections \ref{Sec_trans} and \ref{Sec_block} and installed it into the HAPS and gateway nodes. The NS-3 HAPS module at the HAPS station utilizes the blockchain module to execute the blockchain functions (such as ordering transactions and creating blocks, broadcasting the block and collecting replies, sending “Block ACK” and other messages, etc.). On the other hand, the gateway uses the blockchain module to create transactions from sensor nodes’ data, validate others’ transactions, save them in the pending list, validate new blocks that it receives from HAPS stations, and send voting decisions. Most of the blockchain operations in the blockchain module were based on the source code of Iroha after modifying it to match the required operations in the proposed blockchain model. 

\subsection{Simulation Setup}

After creating the NS-3 modules, we simulated several NS-3 scenarios to test the system performance. We assumed a constellation of ten HAPS stations scattered over an area of 440,896 km\textsuperscript{2} (664 km x 664 km). The size of the network topology was chosen based on the scenarios that we simulated in Section \ref{Sec_eff_haps} to find the effective HAPS footprint when deploying cloud services. The simulation results in Section \ref{Sec_eff_haps} show that in order to achieve an End-to-End delay less than or equal to 100ms, the HAPS station should be serving a maximum of 394 sensor nodes per km\textsuperscript{2} within a total area of 44,000 km\textsuperscript{2}. Based on that, the network size was set to approximately 44,000 x10 km\textsuperscript{2} with ten HAPS stations. Each HAPS station services 394 x 44,000 = 17,336,000 sensor nodes. Hence, the total number of sensor nodes in the simulation scenarios was set to 173,360,000 distributed evenly across the area in order to reach an average of 394 nodes/km\textsuperscript{2}. We assumed that each 100 sensor nodes connect to a single sink (GHG). Hence, we deployed a total of 1,733,600 GHGs that were distributed evenly across the network area. Each sensor node sends a data transaction to the nearest GHG every 100ms. The size of the data transaction was varied between 10 and 1000KB (we tested a separate scenario for each transaction size). To be able to simulate the required scenarios, we used a powerful Core i9 server equipped with 64GB RAM. The remaining simulation parameters are shown in Table \ref{tab:SimPar}.

\begin{table}[!t]
\caption{Simulation parameters
\label{tab:SimPar}}
\centering
\begin{tabular}{|c||c|}
\hline
\textbf{Parameter} & \textbf{Value} \\
\hline
Simulation area &	664 × 664 Km\textsuperscript{2} \\
\hline
Scenario simulation time & 10,000 sec \\
\hline
Number of HAPS stations & 10 \\
\hline
HAPS station altitude & 20km \\
\hline
Number of GHGs & 1,733,600 \\
\hline
Number of sensor nodes & 173,360,000 \\
\hline
Sensor node transmission range & 100m \\
\hline
Data Transaction generation interval & 100ms \\
\hline
Data Transaction size & Varied between 10 - 1000 KB \\
\hline
Block generation interval (\textit{t\textsubscript{th}}) & 100ms \\
\hline
HAPS station waiting period (\textit{t\textsubscript{w}}) & 100ms \\
\hline
Cryptographic algorithms & Ed25519 with SHA-3 \\
\hline
Malicious attack rate & 10s \\
\hline
\% of malicious nodes & 30\% \\
\hline
\end{tabular}
\end{table}

In our experiments, we simulated a direct link between each GHG and the nearest HAPS station. Hence, we didn’t simulate the cases in which AHGs exist. We also didn't simulate the cases in which a cloud service is deployed at specific HAPS stations only, which causes a GHG to send its packets via other GHGs/AHGs if it does not have a direct link with one of these HAPS stations. We plan on extending the simulations in the future to include these cases. 

\color{black}

The adversary model was simulated in the experiments as follows: first, among the seven attacks that were presented in Section \ref{Sec_adversary} and analyzed in Section \ref{Sec_Secanal}, the \textit{Transaction Piracy attack} is automatically resolved with the integration of blockchain. Since each private transaction is encrypted with the receiver's public key, an adversary will not be able to steal private data from the transmitted transaction. In addition, there are three attacks that are resolved by assuming that HAPS stations are trusted nodes and are closely monitored in order to prevent attackers from compromising them. Based on this assumption, the \textit{Block Withholding attack}, the \textit{51\% attack}, and the \textit{Forking and Replay attacks} will never occur, as we explained in Section \ref{Sec_Secanal}. Finally, we assume that the majority of gateways (i.e., GHGs/AHGs) are legitimate, which guarantees that the \textit{Consensus Collusion attack} will not occur. Hence, in this section, we study the remaining two attacks, which are the \textit{False Transaction attack} and the \textit{Consensus Sabotage attack}. \color{black}

In order to test the ability of the proposed system to detect and prevent the two mentioned attacks, we made some of the sensor nodes and GHGs behave maliciously. Hence, a specific percentage of SNs and GHGs were selected as malicious nodes (default value of 30\% of the total number of nodes; however, in a later section we vary this percentage between 10\% and 80\%). A malicious sensor node modifies the simulated sensor readings before sending the packet to the GHG. \color{black}This simulates a \textit{False Transaction attack}. \color{black}If the packet is forwarded by one or more intermediate legitimate sensor nodes, they should detect the malicious modification by comparing the data with their own readings. Hence, these sensor nodes do not endorse malicious data when they forward the packet to the next sensor node or to the GHG. When the GHG receives a packet that was not endorsed by one or more sensor nodes, it checks the data in the packet by comparing it to the data that was generated by the near sensor nodes at similar times. If the GHG finds out that the data values are majorly different from the values that it received from the near sensor nodes, it discards the packet. On the other hand, if the sensor node is a neighbor to the GHG, it sends its packets directly to it. If a GHG receives a packet that was not endorsed by an intermediate sensor node, it performs the same action described before (i.e., comparing the data with that generated by sensor nodes that are near the sender). If the GHG still does not have the readings from the near sensor nodes, it waits until it receives them. 

With respect to malicious GHGs, we simulated their attacks as follows: a malicious GHG attempts to affect the consensus operation by sending a “Block ERROR” message to the HAPS station to increase the consensus delay. \color{black}This simulates a \textit{Consensus Sabotage attack}. \color{black}If the HAPS station receives “Block ERROR” messages from a few GHGs but receives “Block ACK” from all the HAPS stations and the remaining GHGs, it revalidates the transactions that were marked as erroneous by the GHGs. If the HAPS station finds out that these transactions are valid, it discards the “Block ERROR” messages and generates a warning report. In the simulations, we created an administrative program that responds to the warning report from the HAPS station by stopping the execution of the code that leads to the malicious behavior by these GHGs and switching them back to normal behavior, which simulates fixing these GHGs. In order to maintain the percentage of malicious nodes, the program selects other legitimate GHGs and switches them to malicious mode. Our aim is to find out the percentage of cases in which a malicious GHG will be detected and the effect of the GHGs' malicious behavior on the consensus delay, which we will discuss in the simulation results. Note that if the HAPS station receives “Block ERROR” from more than half of the GHGs, it does not approve the new block directly. Rather, it generates the warning report as mentioned before and waits for it to fix the malicious GHGs, then it resends the new block to these GHGs and waits for their replies before it reevaluates the voting results. This was done according to the QUICO consensus condition which was explained in Section \ref{Sec_block}. 

To test the effectiveness of the proposed system, we compared its results with those of the system proposed in \cite{fu2019resource}. The compared system, MOMEC, utilizes the blockchain to facilitate autonomous management and orchestration of virtualized resources in a Mobile Edge Cloud. In addition, MOMEC proposes a custom consensus protocol to secure cloud services that are offered by mobile edge servers. When we simulated MOMEC within the HAPS environment, we made the HAPS stations act as the NFV MANO servers, while the ground gateways acted as the MEC Servers. The two compared systems were evaluated based on the following parameters: 1) Blockchain Throughput (\textbf{BTh}), which is the number of transactions added to the blockchain per second, 2) Transaction Latency (\textbf{TLa}), which is the time between the transaction is sent by the sensor node and the block that contains the transaction is confirmed by the consensus protocol, 3) Consensus Time (\textbf{CT}), which is the time between the instance a new block is broadcast by the HAPS station until the instance the block is added to the blockchain at all HAPS stations, 4) Attack Detection Rate (\textbf{ADR}), which is the percentage of malicious transactions that are detected by the HAPS stations, 5) Malicious GHG Detection Rate (\textbf{MGDR}), which is the percentage of times a malicious GHG action is detected by the HAPS stations, and 6) Network Traffic (\textbf{NT}), which is the average traffic sent, forwarded, and received by a node (sensor and GHG) per second.

\subsection{Simulation Results}

\subsubsection{Effective HAPS Footprint}
\label{Sec_eff_haps}

In this section, we present the simulations that we performed to determine the effective HAPS footprint when offering cloud service to ground users. Note that previous HAPS papers, such as \cite{kurt2021vision}, assumed the HAPS footprint as a ground circle with a radius of 500 km, but this estimate did not consider that the HAPS will be deploying cloud services. Rather, previous systems considered the HAPS as a super macro base station (SMBS) or as an aerial gateway for backhauling isolated BSs. When deploying and offering cloud services, the HAPS will experience a high load from ground users that will limit its capability to cover a wide footprint while satisfying the QoS requirements of the cloud service. For this reason, we tested a set of four simulation scenarios in which the network topology was set to (50 km x 50 km = 2500 km\textsuperscript{2}), (100 km x 100 km = 10,000 km\textsuperscript{2}), (200 km x 200 km = 40,000 km\textsuperscript{2}), and (400 km x 400 km = 160,000 km\textsuperscript{2}). We will refer to these scenarios as Map\textsubscript{1}, Map\textsubscript{2}, Map\textsubscript{3}, and Map\textsubscript{4}. Next, we obtained the average population in urban areas from \cite{bureau_2021}, the average percentage of people who use the Internet from \cite{facts_2021}, and the percentage of Internet users who use cloud services from \cite{cso_2020}. Based on that, we found that the average cloud user density is equal to 394.15 individuals per km\textsuperscript{2}.

From these calculations, the total number of cloud users that were created in the four simulation scenarios was set to (2,500 x 394.15 = 985,375), (10,000 x 394.15 = 3,941,500), (40,000 x 394.15 = 15,766,000), and (160,000 x 394.15 = 63,064,000) users, respectively. In each scenario, the users were scattered uniformly within the network topology. A single HAPS was placed in the center of the topology. Finally, several GHGs were placed in each scenario such that each GHG serves 100 users. Each cloud user sends its requests to the nearest GHG, and the latter forwards them to the HAPS station and vice versa. 

\begin{figure}[!t]
\centering
\includegraphics[width=4.5in]{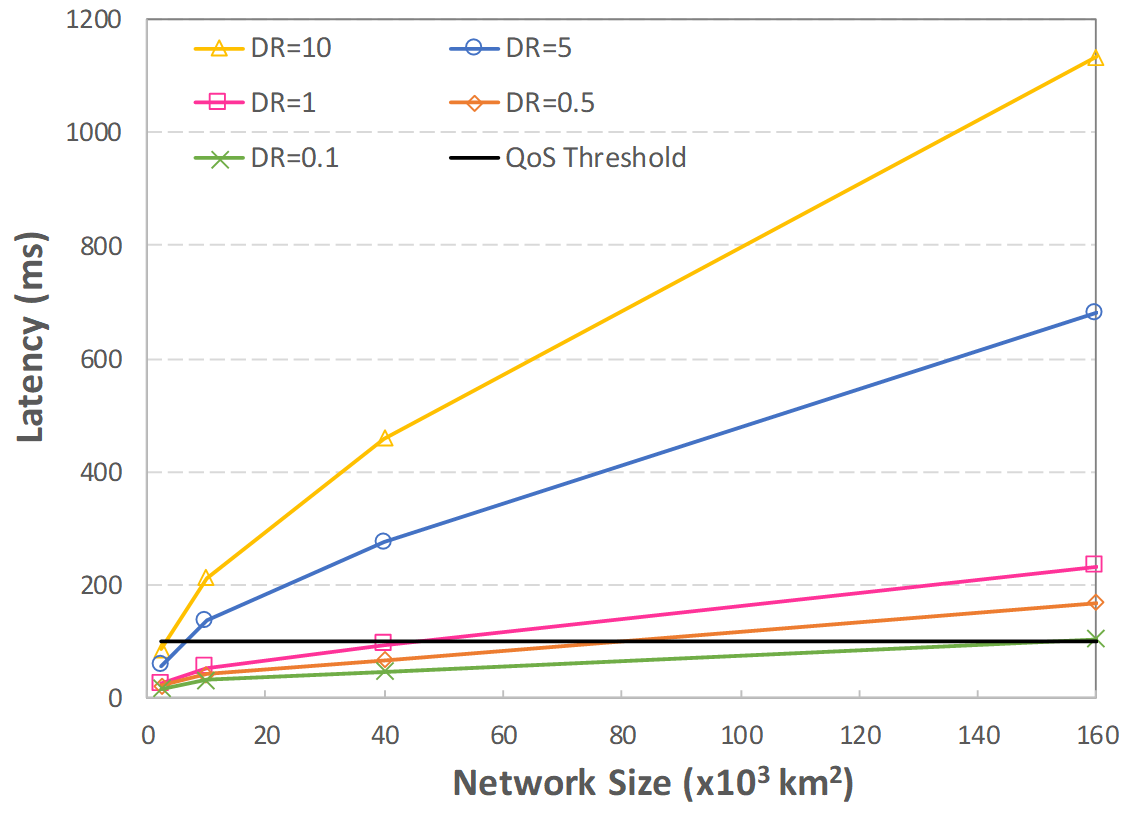}
\caption{The average latency for various data rates in Map\textsubscript{1}, Map\textsubscript{2}, Map\textsubscript{3}, and Map\textsubscript{4}.}
\label{fig_Res1}
\end{figure}

The users' data request rate was varied between 100 Kbps and 10 Mbps. For this purpose, we simulated five scenarios in each of the four maps. The data rate in the five scenarios was set to 100Kbps, 500Kbps, 1Mbps, 5Mbps, and 10Mbps. In order to set the user data rate, we made each node send a request to the HAPS each 100 ms and the HAPS sends back a reply packet. The size of the request packet was set to 128 bits while the size of the reply packet was varied in each scenario to achieve the required data rate. Hence, the size of the reply packet was set to 10, 50, 100, 500, and 1000 Kbits in the five scenarios. In each scenario, we calculate the average End-to-End latency from the instance a cloud user sends a data request packet until it receives the HAPS station reply. The results are shown in Fig. \ref{fig_Res1}.

As stated before, different cloud services require different delay thresholds to achieve a good quality of service (QoS). For example, cloud gaming requires a maximum latency of 20ms. In the simulated scenarios, we consider a latency of 100ms as the maximum delay threshold. In other words, when the request delay is greater than 100ms, we consider that the HAPS station failed to provide the cloud service to the user with accepted QoS. Our aim is to study the maximum number of users that the HAPS station is able to serve with accepted QoS for each of the 5 data rates, and hence the HAPS effective footprint. From Fig. \ref{fig_Res1}, we deduce that the effective HAPS footprint is equal to 150,000 km\textsuperscript{2}, 78,000 km\textsuperscript{2}, 44,000 km\textsuperscript{2}, 7,000 km\textsuperscript{2}, and 3,000 km\textsuperscript{2} for the five data rates. If we consider an average data rate of 1 Mbps, then the effective HAPS footprint will be 44,000 km\textsuperscript{2}, which is the value that was used to calculate the total network size in the simulations (i.e., 10 HAPS stations x 44,000 km\textsuperscript{2} per station).

\begin{figure}[!t]
\centering
\includegraphics[width=4.5in]{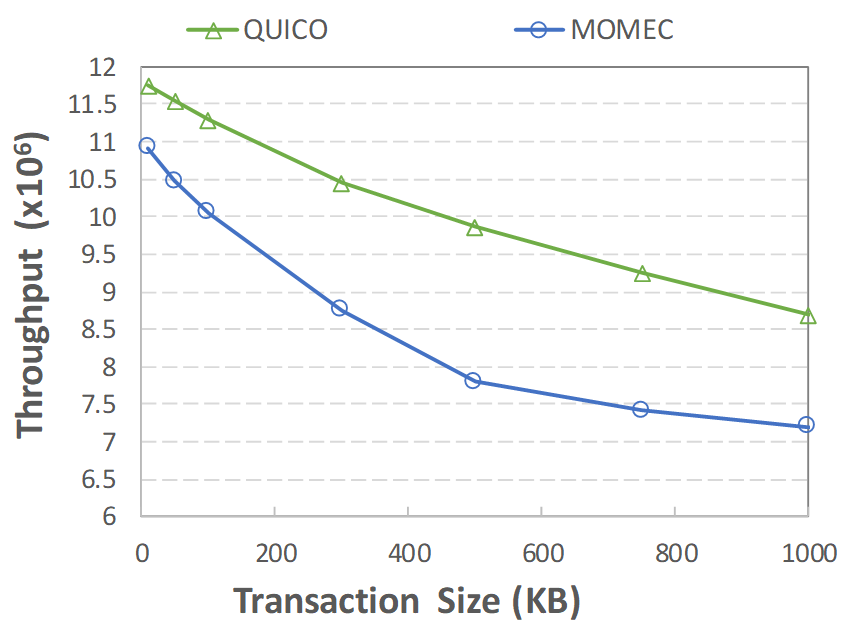}
\caption{Blockchain throughput of QUICO and MOMEC while varying the transaction size.}
\label{fig_Res2}
\end{figure}

\subsubsection{Varying the Transaction Size}
\label{Sec_Var_tran_size}

\begin{figure}[!t]
\centering
\includegraphics[width=4.5in]{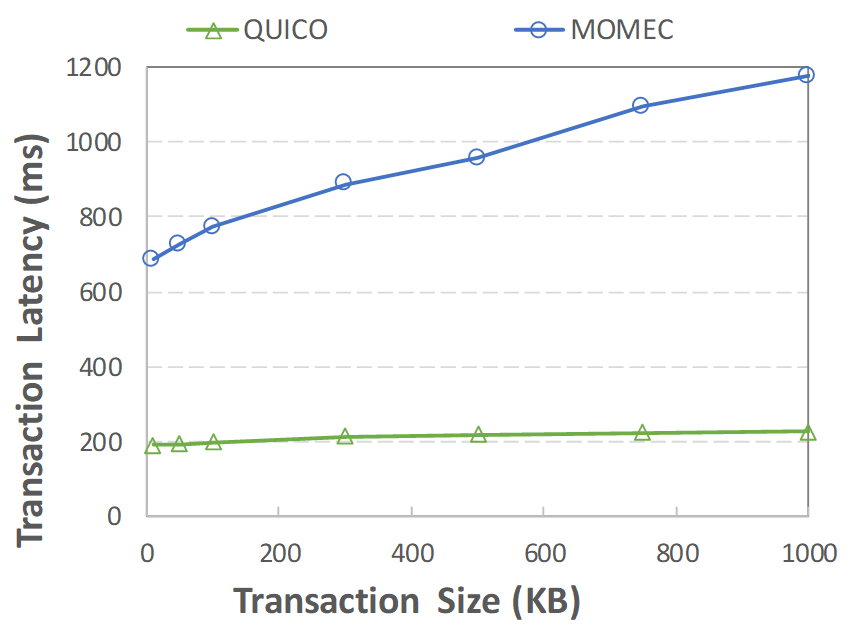}
\caption{Transaction latency of QUICO and MOMEC while varying the transaction size.}
\label{fig_Res3}
\end{figure}

After calculating the HAPS station's effective footprint, we fix the user data rate at 1Mbps while varying the sensor node transaction size between 10 and 1000KB in the next set of simulation scenarios. Fig. \ref{fig_Res2} shows the blockchain throughput of the two compared systems. We notice that the throughput generally decreases as the transaction size increases. This is due to the fact that a larger-size transaction is divided into a larger number of packets when being transmitted (due to the MTU limit). In general, this increases the delay for the transaction to reach its destination. When this delay becomes greater than the HAPS station waiting time (100ms in the simulations), the transaction is not added to the current block but remains in the pending list until the next block is generated. Hence, when the size of the transaction increases, the probability that the total delay until the destination receives the last packet of the transaction is greater than 100ms increases. This means that more transactions are delayed to later blocks, which reduces the number of transactions in the block and decreases the throughput. Fig. \ref{fig_Res2} shows that QUICO has a higher throughput than MOMEC (by an average of 12\%) which is mainly due to the fact that QUICO has a much lower transaction latency, on average, as shown in Fig. \ref{fig_Res3}. 

\color{black}
Table \ref{tab:Confid1} illustrates the confidence in the results of Fig. \ref{fig_Res2}. The confidence was calculated by taking a sample of 1M readings from each scenario. Table \ref{tab:Confid1} shows the confidence values for four different significance levels (0.05, 0.1, 0.2, and 0.3), which correspond to 95\%, 90\%, 80\%, and 70\% confidence levels, respectively. From the table, the confidence interval becomes smaller as the significance level decreases, which is logical. The maximum confidence value is 0.715 which occurs when the average value is equal to 10.453 x 10\textsuperscript{6} and the significance level is 0.05. At this value, the confidence interval is equal to [9.74, 11.17] x 10\textsuperscript{6} with a 95\% confidence level. Such confidence interval is fairly acceptable, taking into consideration the huge sample size and the variations in the network conditions. In general, the confidence values in the table reflect an overall high accuracy in the simulation results, since the confidence values show that there is no high deviation from the average value in all the confidence intervals.
\color{black}

\color{black}
\begin{table}[!t]
\begin{adjustwidth}{-1.5cm}{}
\caption{Confidence levels in the results of Fig. \ref{fig_Res2} for different values of the significance level. 
\label{tab:Confid1}}
\centering 
\begin{tabular}{|p {2.5 cm}|p {1.8 cm}|p {2.6 cm}|p {2.5 cm}|p {2.5 cm}|p {2.5 cm}|}
\hline 
\textbf{Transaction Size (KB)} & \textbf{Average BTh} & \textbf{Significance Level = 0.05} & \textbf{Significance Level = 0.1} & \textbf{Significance Level = 0.2} & \textbf{Significance Level = 0.3}\\ \hline
10 & 11.750 & 0.687 & 0.576 & 0.454 & 0.354 \\ \hline 
50 & 11.550 & 0.707 & 0.582 & 0.466 & 0.368 \\ \hline 
100 & 11.284 & 0.713 & 0.587 & 0.470 & 0.371 \\ \hline 
300 & 10.453 & 0.715 & 0.595 & 0.452 & 0.365 \\ \hline
500 & 9.859 & 0.704 & 0.584 & 0.433 & 0.357 \\ \hline
750 & 9.252 & 0.681 & 0.562 & 0.412 & 0.345 \\ \hline
1000 & 8.688 & 0.653 & 0.538 & 0.393 & 0.333 \\ \hline
\end{tabular}
\end{adjustwidth}
\end{table}

\color{black}

In Fig. \ref{fig_Res3}, the transaction latency is measured from the instance the transaction owner (i.e., sensor node) sends the transaction until the instance the transaction is added to the blockchain. From Fig. \ref{fig_Res3}, the average TLa of QUICO varies between 189 and 226ms while that of MOMEC varies between 686ms and 1.17s. This huge difference is due to two main reasons: first, the average consensus delay of QUICO is much smaller, as we will explain soon, and second, the block rejection method in MOMEC (which is adopted from traditional PBFT) is replaced in QUICO with the block check approach (i.e., using the “ERROR Check” and “ERROR Resolve” packets). In MOMEC, if consensus is not reached over a certain block, COMMIT is not made and the block is rejected, which highly increases the TLa of the block transactions since they will wait for the next block to be added to the blockchain. In QUICO, the HAPS station does not reject a block but resolves the erroneous transactions as explained before. This highly decreases the average delay of transactions.

\begin{figure}[!t]
\centering
\includegraphics[width=4.5in]{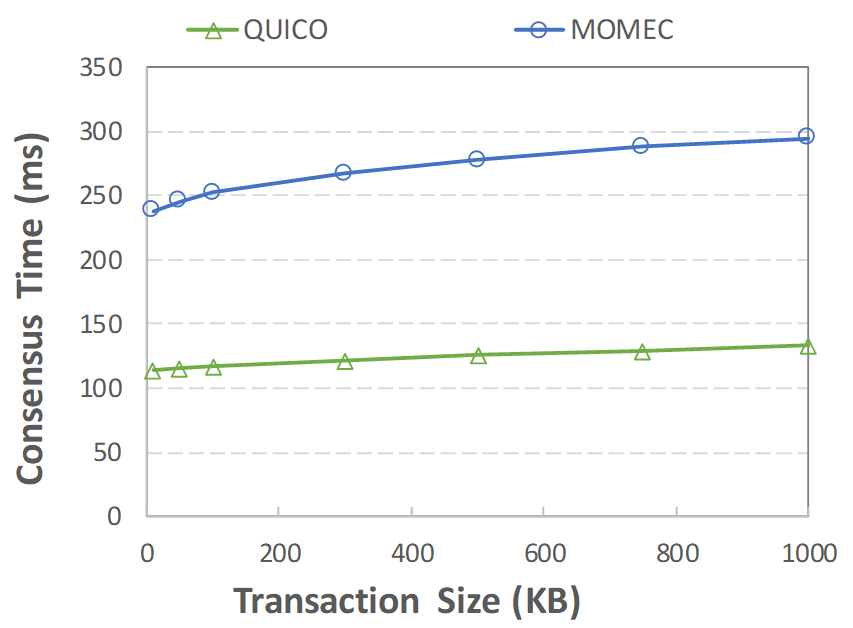}
\caption{Consensus delay of QUICO and MOMEC while varying the transaction size.}
\label{fig_Res4}
\end{figure}

The consensus delay is shown in Fig. \ref{fig_Res4}. This parameter is one of the major advantages of QUICO as compared to MOMEC and traditional PBFT. Instead of broadcasting PRE-PREPARE and COMMIT messages between the consensus nodes, QUICO relies on the trustworthiness of HAPS stations to reduce the communication and delay overheads. Hence, the new block is broadcast to all consensus nodes which reply to the block creator directly and the consensus decision is made by the latter. The process of avoiding the broadcast storm in the COMMIT phase reduces the consensus delay by approximately 50\%, as shown in Fig. \ref{fig_Res4}. The CT of MOMEC is higher than that of QUICO by 109\% when the transaction size is equal to 10KB and by 121\% when the latter is equal to 1000KB.

\begin{figure}[!t]
\centering
\includegraphics[width=4.5in]{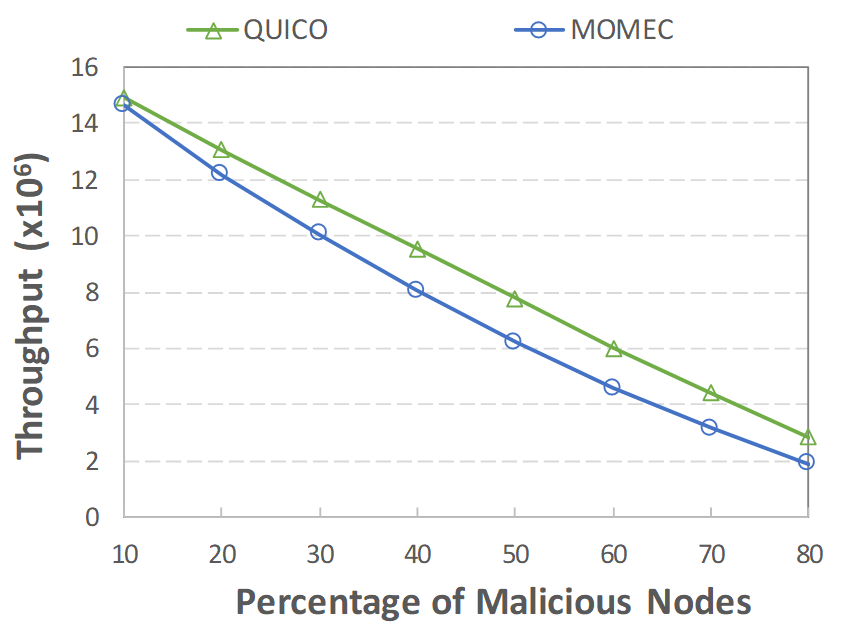}
\caption{Blockchain throughput of QUICO and MOMEC while varying the percentage of malicious nodes.}
\label{fig_Res5}
\end{figure}

\subsubsection{Varying the Percentage of Malicious Nodes}
\label{Sec_Var_mal}

In the next set of simulations, we vary the percentage of malicious nodes (both sensor nodes and GHGs) between 10 and 80\%. Fig. \ref{fig_Res5} shows that the throughputs of the two systems decrease in a similar fashion as the percentage of malicious nodes (PMN) increases. This is logical since when the number of malicious sensor nodes increases, the number of malicious transactions increases, and hence, the number of valid transactions decreases. Since the two systems have high attack detection rates, as we will discuss soon, most of the malicious transactions are detected by the legitimate GHGs and the HAPS stations and will not be added to the blockchain. This causes a decrease in the number of transactions that are added to the blockchain and hence a lower throughput.

\begin{figure}[!t]
\centering
\includegraphics[width=4.5in]{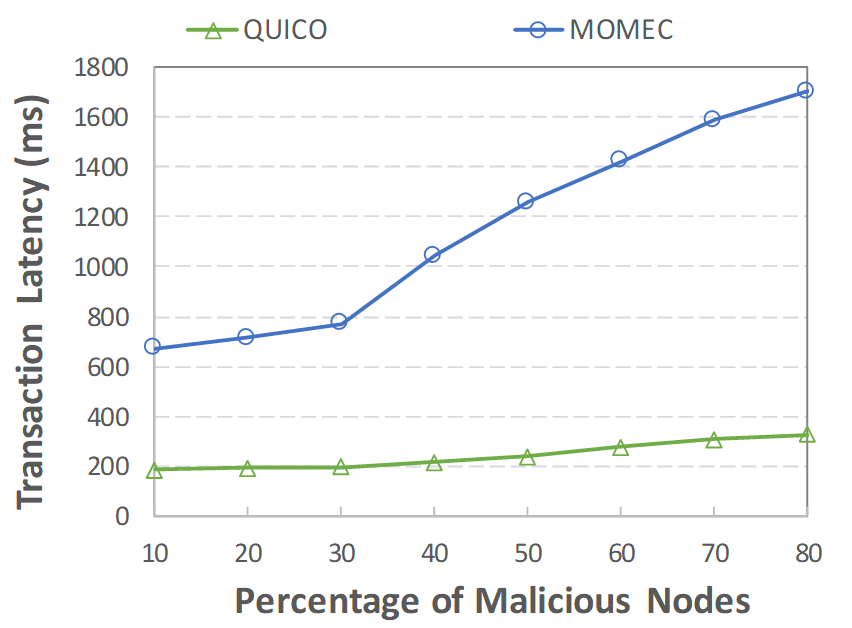}
\caption{Transaction latency of QUICO and MOMEC while varying the percentage of malicious nodes.}
\label{fig_Res6}
\end{figure}

With respect to the transaction latency, we notice from Fig. \ref{fig_Res6} that the TLa of MOMEC is much more affected by the increase in PMN than that of QUICO. We notice that this high increase in the Tla of MOMEC occurs when PMN is higher than 30\%. This is due to the consensus property of MOMEC and PBFT which requires at least two-thirds of the consensus nodes to COMMIT in order to add the block. If more than 1/3 of the nodes do not COMMIT the block, then the latter is rejected by the consensus nodes. This explains the high surge in the TLa of MOMEC when PMN is 40\% or higher. In such cases, most of the blocks are rejected and the transactions should wait for the next block to be added to the blockchain, which highly increases their latency. On the other hand, a smaller surge occurs in the TLa of QUICO when PMN increases above 50\%, since in such case, the HAPS station will receive “BLOCK ACK” from less than half the GHGs and will wait until the malicious GHGs are fixed and they send “BLOCK ACK” before the HAPS station commits the block. This wait causes a small rise in the TLa of QUCIO when PMN is greater than 50\%, as can be seen in Fig. \ref{fig_Res6}.

\begin{figure}[!t]
\centering
\includegraphics[width=4.5in]{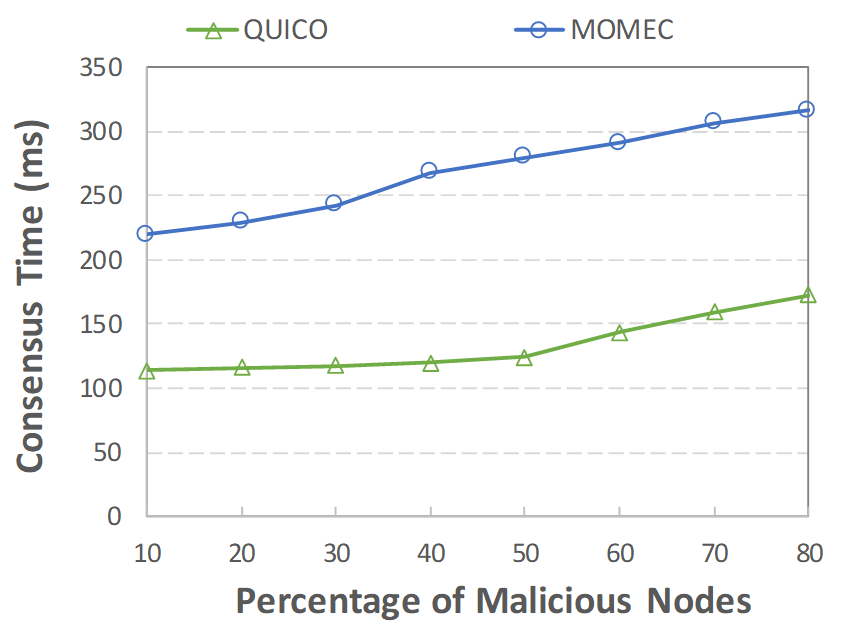}
\caption{Consensus delay of QUICO and MOMEC while varying the percentage of malicious nodes.}
\label{fig_Res7}
\end{figure}

A similar observation can be noted about the consensus time in Fig. \ref{fig_Res7}. The CT of both systems increases with PMN, however; the surge in the CT of MOMEC occurs after PMN = 30\%, while that of QUICO happens when PMN is greater than 50\%. Note that the CT is also affected by the increase in the waiting time of the HAPS station. When the percentage of malicious nodes is less than the consensus threshold (50\% for QUICO and 30\% for MOMEC), the HAPS station has a higher probability of receiving the COMMIT messages from the required number of consensus nodes and hence reaching consensus earlier. On the other hand, when the percentage of malicious nodes is greater than the consensus threshold, the HAPS station will wait until it receives replies from most of the consensus nodes to make sure that the block has been accepted or rejected by the majority of nodes (i.e., more than 50\% in QUICO and 60\% in MOMEC) before it takes the consensus decision. This increases the average consensus delay.

\begin{figure}[!t]
\centering
\includegraphics[width=4.5in]{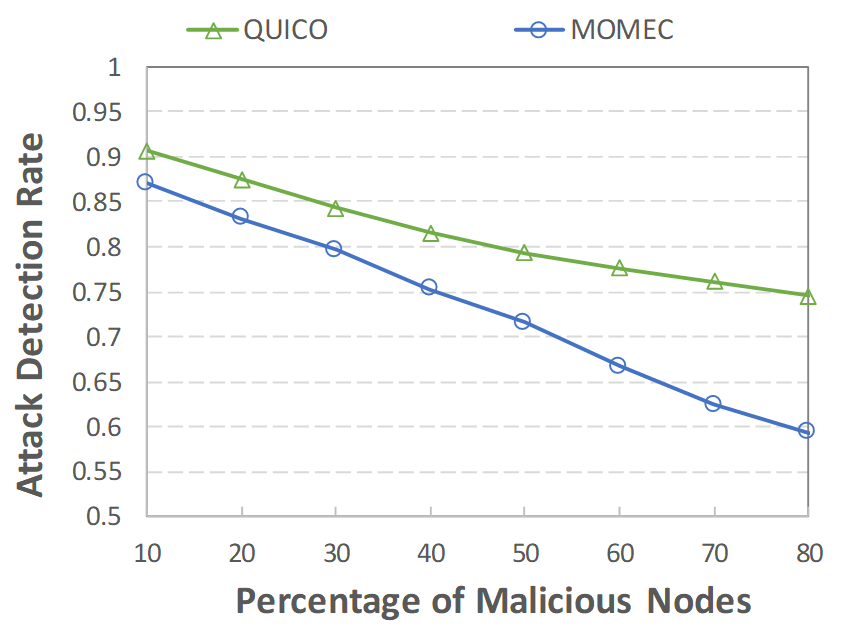}
\caption{Attack detection rate of QUICO and MOMEC while varying the percentage of malicious nodes.}
\label{fig_Res8}
\end{figure}

\subsubsection{Attack and GHG Detection Rates}
\label{Sec_det_rate}

While varying PMN between 10 and 80\%, we measured the sensor nodes’ and GHG attack detection rate. With respect to the ADR, we added a secret parameter to each transaction which indicates whether it is legitimate or malicious (this parameter is secret because it was not read by network nodes but used when calculating the results only). Next, we calculated the number of malicious transactions that were added to each new block and divided it by the number of malicious transactions that were generated by the sensor nodes. Finally, we took the average over all blocks and subtracted it from 100\% to calculate the ADR. Fig. \ref{fig_Res8} shows that both QUICO and MOMEC have acceptable ADRs when PMN is small. However, the ADR of MOMEC decreases to 59\% when PMN is 80\% while that of QUICO decreases to 75\% only. The main reason is that QUICO adopts the endorsement strategy which helps legitimate GHGS and HAPS stations detect malicious transactions that were not endorsed by sensor nodes that forwarded them. In addition, QUICO makes the GHGs and HAPS stations validate the sensor’s data by comparing them to data that was sent by other near sensors at a similar time. A big difference in the data could indicate a malicious transaction. These strategies contribute to increasing the ADR of QUICO. On the other hand, MOMEC implements the default verification model of PBFT. \color{black}Table \ref{tab:Confid2} illustrates the confidence in the results of Fig. \ref{fig_Res8}. Similar to the results in Table \ref{tab:Confid1}, the confidence in the ADR results is high, with the largest confidence interval equal to [0.69, 0.75] when PMN is 80\%.
\color{black}

\begin{table}[!t]
\begin{adjustwidth}{-1.5cm}{}
\caption{Confidence levels in the results of Fig. \ref{fig_Res8} for different values of the significance level. 
\label{tab:Confid2}}
\centering 
\begin{tabular}{|p {2.5 cm}|p {1.8 cm}|p {2.6 cm}|p {2.5 cm}|p {2.5 cm}|p {2.5 cm}|}
\hline
\textbf{Percentage of Malicious Nodes} & \textbf{Average ADR} & \textbf{Significance Level = 0.05} & \textbf{Significance Level = 0.1} & \textbf{Significance Level = 0.2} & \textbf{Significance Level = 0.3}\\
\hline
10 & 0.907 & 0.0283 & 0.0226 & 0.0174 & 0.0139
 \\ \hline
20 & 0.875 & 0.0322 & 0.0245 & 0.0190 & 0.0186
 \\ \hline 
30 & 0.843 & 0.0330 & 0.0272 & 0.0215 & 0.0218
 \\ \hline 
40 & 0.815 & 0.0383 & 0.0303 & 0.0242 & 0.0233
 \\ \hline 
50 & 0.792 & 0.0406 & 0.0331 & 0.0273 & 0.0238
 \\ \hline 
60 & 0.776 & 0.0422 & 0.0380 & 0.0230 & 0.0254
 \\ \hline 
70 & 0.761 & 0.0469 & 0.0411 & 0.0309 & 0.0272
 \\ \hline 
80 & 0.745 & 0.0505 & 0.0415 & 0.0321 & 0.0273
 \\ \hline 
\end{tabular}
\end{adjustwidth}
\end{table}

\begin{figure}[!t]
\centering
\includegraphics[width=4.5in]{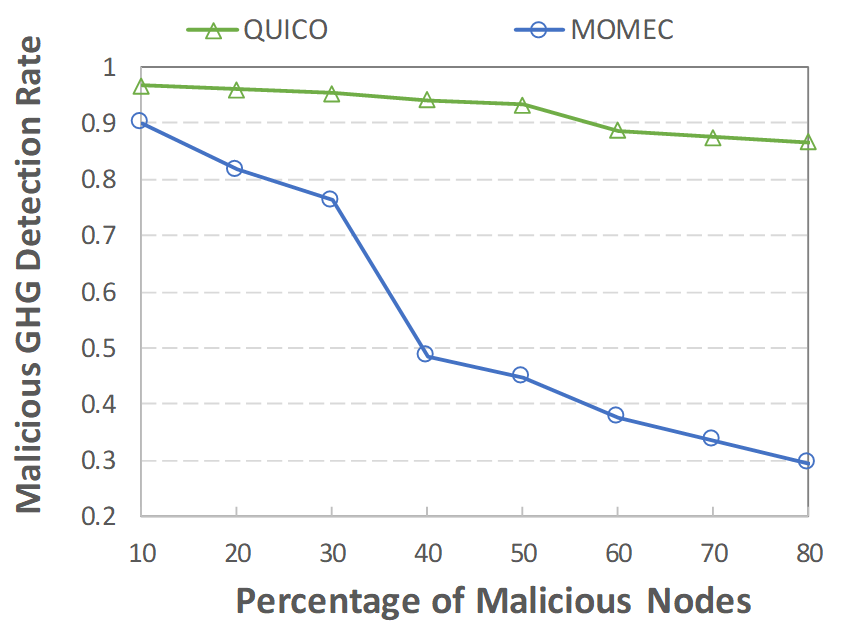}
\caption{Malicious GHG detection rate of QUICO and MOMEC while varying the percentage of malicious nodes.}
\label{fig_Res9}
\end{figure}

With respect to the malicious GHG detection rate, Fig. \ref{fig_Res9} shows that QUICO has a much higher capability of detecting a malicious GHG than MOMEC, especially at a high PMN. QUICO applies a clear strategy to detect a GHG’s malicious behavior. As explained in Section \ref{Sec_block}, a HAPS station checks with the GHGs who sent the transactions that were labeled as erroneous by the malicious GHGs to validate them (via “ERROR Check” packets). In addition, the HAPS station generates a warning report to the network administrators when it detects a possible malicious GHG. The methods applied in QUICO enable HAPS stations to detect malicious GHGs with a high probability (between 86 and 97\%, as shown in Fig. \ref{fig_Res9}). On the other hand, MOMEC does not adopt any method to detect malicious consensus nodes, which causes its MGDR to drop quickly as the percentage of malicious GHGs increases above 30\%. In cases where the majority of GHGs are malicious, consensus will not be reached and the HAPS station will not be able to detect which GHGs are malicious. The authors of \cite{fu2019resource} did not discuss how to detect a malicious consensus node in their paper. Hence, the HAPS stations will consider any GHG that made a decision different from the consensus as malicious, which is a false deduction in many cases.

\color{black}

\subsubsection{SN Energy Consumption}
\label{Sec_energy}
One of the main challenges of integrating the blockchain into a WSN is related to dealing with the blockchain overhead on SNs. In this section, we study the effect of the proposed blockchain model on the SN energy consumption. For this purpose, we integrated the NS-3 Energy Framework that was published in \cite{wu2012energy} into our simulations. We modified the parameters of the “Energy Source” and “Device Energy Model” NS-3 classes (we replaced some parameters and added other parameters) to include the factors related to the blockchain operations. Our modifications were based on the Lr-WPAN-MAC model that was proposed in \cite{rege2016realistic} and obtained from the lr-wpan code repository\footnote[1]{https://code.nsnam.org/vrege/ns-3-gsoc/file/c43c335bc921/src/lr-wpan/model}. In the simulation scenarios, We initiate the Energy Source of each SN to have a total energy density equal to 0.2775Wh (watt-hour) at the beginning of each simulation scenario, which is equivalent to 1000 Joules. This parameter denotes the maximum amount of energy that an SN can consume during the simulation scenario. This value was obtained by simulating the SN to have a battery capacity of 75mAh and 3.7V, which can be considered as average battery specifications for a general sensor node. 
 
\begin{figure*}[!t]
\centering 
\includegraphics[width=4.5in]{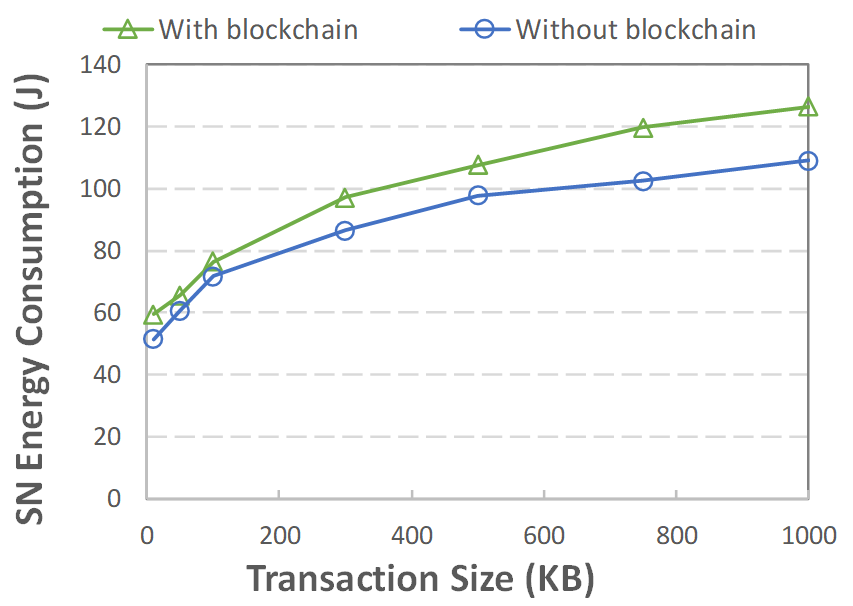} 
\caption{\color{black}Total energy consumption during the simulation time per SN, with and without the blockchain, while varying the transaction size.\color{black}} 
\label{fig_Res10} 
\end{figure*}

In this section, we simulated two main types of scenarios: the first is the same as the simulation scenarios that we presented in Section \ref{Sec_Var_tran_size}, while the second scenario does not include the proposed blockchain model. Rather, in the second scenario, SNs send their readings to the CHs who forward them to GHGs/AHGs, and the latter aggregate the readings into data transactions and send them to the HAPS stations which store them in their storage units. In addition, cloud users send their requests to the GHGs/AHGs and the latter forward them to the HAPS stations who reply to the clients via the gateways. Both scenarios were simulated while varying the Transaction size between 10 and 1000KB. For each scenario, we calculate the SN total energy consumption during the simulation time and take the average for all SNs. The results of the two scenarios are shown in Fig. \ref{fig_Res10}.

From Fig. \ref{fig_Res10}, we can see that the blockchain contributes to a slight increase in the SN total energy consumption. As stated before, the SN does not directly participate in the blockchain operations, such as creating and storing blocks and executing the consensus protocol. These operations are performed by the GHGs/AHGs and HAPS stations only. On the other hand, the SN validates and endorses the data of other SNs and sends/forwards data packets to the CH. In addition, the SN could cache its data for a specific period. In such case, the SN would be contacted by the GHG to resolve questionable transactions, as explained in Section \ref{Sec_block}. Note that these operations are performed by the SN with and without the blockchain integration. Hence, the latter should not have a direct effect on the SN resources. Rather, the slight increase in the SN's energy consumption is due to the effect of the blockchain on the whole network, which indirectly affects the SN. For example, the blockchain (especially the consensus protocol) requires a lot of network communications and broadcasting which congests the network and increases the percentage of resent packets due to intermediate nodes (such as CHs and GHG) dropping some packets. 
This observation can be deduced from the figure: as the transaction size increases, a larger number of packets is generated (due to packet segmentation), which increases the congestion. Hence, the number of dropped/resent packets increases, and consequently, the energy consumption increases. However, the increase in the SN energy consumption is limited (15.6\% when the transaction size is equal to 1000KB).

\color{black}

\subsubsection{Measuring the Network Traffic}
\label{Sec_traff}

\begin{figure*}[!t]
\centering
\includegraphics[width=5.5in]{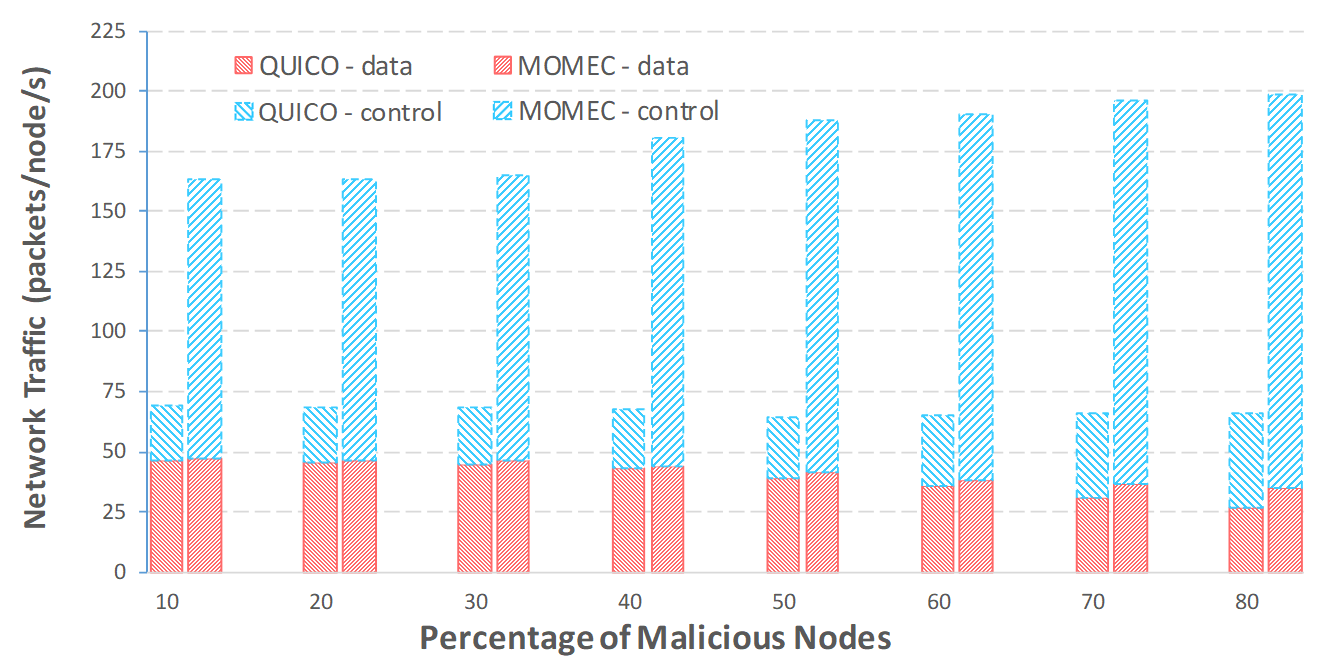}
\caption{Network traffic per node per second of QUICO and MOMEC while varying the percentage of malicious nodes.}
\label{fig_Res11}
\end{figure*}

The last parameter that we calculated is the network traffic. While varying the PMN, we defined six parameters in each scenario: three for data packets and three for control packets. The three parameters are the total number of packets sent, received, and forwarded respectively. Each time a sensor node or GHG creates a new data packet and sends it, we increment the Nb\_data\_sent parameter. Similarly, each time a sensor node or GHG receives a data packet, processes it (for example, endorses the data), and then forwards it, we increment the Nb\_data\_forward parameter. Finally, each time a GHG receives a data packet from a sensor node, we increment the Nb\_data\_recv parameter. We did the same thing for control packets, which include the various consensus packets required by each protocol. Note that QUICO and MOMEC use different control packets in the consensus protocol as explained before. At the end of each scenario, we add the three parameters of each type and divide by the total number of nodes (sensors and GHGs) and by the total simulation time to find the NT in packet per (node second). Fig. \ref{fig_Res11} illustrates that the two systems produce a similar number of data packets on average. On the other hand, MOMEC generates much more control packets than QUICO. This is mainly due to two reasons: 1) the consensus process in MOMEC contains two broadcast stages (PRE-PREPARE and COMMIT) while that in QUICO contains a single broadcast stage, and 2) When a block is rejected in QUICO, another block should be created instead and the consensus process is repeated, which doubles the number of consensus packets required to commit the block. Fig. \ref{fig_Res11} shows that the number of control packets in MOMEC is higher by about 405\% on average than that in QUICO.

\section{\color{black} {Conclusions and Future Works}}
\label{Sec_conc}

\color{black} {HAPS provide a great opportunity for cloud providers to enhance the quality of their services and expand their outreach. Securing C-HAPS platforms, however, becomes vital, as cloud services are susceptible to a wide range of cyberattacks that target their software, data, and network connections. This paper proposes a blockchain model as a solution to secure the C-HAPS EIM application from major cloud-related attacks. The paper describes the network architecture of the studied application and the blockchain role of each node in the system. The paper presents the details of the system implementation, including the storing and consuming of cloud transactions, the generation of new blocks, and the blockchain consensus protocol specifically design to account to the EIM requirements. The paper results, particularly, illustrate the performance of the proposed system in terms of throughput, latency, and resilience to attacks, which highlights the importance of the proposed model in securing future C-HAPS applications.

 Future works in the field include integrating machine learning capabilities into the proposed blockchain operations such as smart contract functions and consensus. The idea herein is that, instead of applying a direct formula to reach consensus, a machine learning algorithm can be applied to evaluate the actions of gateways and select specific gateways to participate in the consensus. Other potential open issues are related to improving the accuracy of the proposed consensus mechanism by assigning validation tasks to sensor nodes in a secure manner. Another possible enhancement is to divide the block validation process among the sensor nodes in the cluster so as to validate very large blocks. Other future research directions is to further investigate the optimal deployment of HAPS stations and gateways to reduce the transaction and consensus delays of the blockchain while keeping the traffic overhead within acceptable limits. Such a direction can also be considered with a joint routing protocol for reliably routing the data packets from the source GHG to the destination HAPS station so as to reduce the end-to-end latency promises, especially given the ever-growing interest in massive ultra-reliable low latency communications (mURLLC) from the sky.}


 \bibliographystyle{elsarticle-num} 
 \bibliography{main}





\end{document}